\documentclass{article}

\usepackage{arxiv}

\usepackage[utf8]{inputenc} 
\usepackage[T1]{fontenc}    
\usepackage{hyperref}       
\usepackage{url}            
\usepackage{booktabs}       
\usepackage{amsfonts}       
\usepackage{nicefrac}       
\usepackage{microtype}      
\usepackage{lipsum}
\usepackage[square,numbers]{natbib}
\usepackage{listings}
\usepackage{amsmath}
\usepackage{amssymb}
\usepackage{ulem}
\usepackage{caption}
\usepackage{subcaption}
\usepackage{wrapfig}
\usepackage{tikz}
\usetikzlibrary{shapes,arrows}
\usepackage{makecell}
\usepackage{multirow}

\usepackage[]{hyperref}

\usepackage[colorinlistoftodos,prependcaption, textsize=tiny,textwidth=2cm]{todonotes}

\newcommand{\fancyseir}{$\text{SEI}^3\text{R}$}
\newcommand{\cov}{COVID-19}
\usepackage{authblk}

\title{Planning as Inference \\in Epidemiological Dynamics Models}

\date{}

\author[1,3,4]{\small Frank Wood}
\author[2]{\small Andrew Warrington}
\author[1]{\small Saeid Naderiparizi}
\author[1]{\small Christian Weilbach}
\author[1]{\small Vaden Masrani}
\author[1]{\\ \small William Harvey}
\author[1]{\small Adam \'Scibior}
\author[1]{\small Boyan Beronov}
\author[5]{\small John Grefenstette}
\author[5]{\small Duncan Campbell}
\author[1]{\small Ali Nasseri}

\affil[1]{\footnotesize Department of Computer Science, University of British Columbia}
\affil[2]{\footnotesize Department of Engineering Science, University of Oxford}
\affil[3]{\footnotesize MILA}
\affil[4]{CIFAR AI Chair}
\affil[5]{Epistemix Inc., Pittsburgh}
\affil[ ]{\textit
{\{fwood,awarring,saeidnp,weilbach,vadmas,wsgh,ascibior,beronov\}@cs.ubc.ca,
ali.nasseri@ubc.ca,
\{john.grefenstette,duncan.campbell\}@epistemix.com}}

\newcommand{\changed}[3][None]{#3}

\renewcommand{\emph}{\textit}

\begin{document}
\maketitle
\begin{abstract}
\changed{In this work, we demonstrate how existing software tools can be used to automate parts of infectious disease-control policy-making via performing inference in existing epidemiological dynamics models. The kind of inference tasks undertaken include computing, for planning purposes, the  posterior distribution over putatively controllable, via direct policy-making choices, simulation model parameters that give rise to acceptable disease progression outcomes. Neither the full capabilities of such inference automation software tools nor their utility for planning is widely disseminated at the current time. Timely gains in understanding about these tools and how they can be used may lead to more fine-grained and less economically damaging policy prescriptions, particularly during the current COVID-19 pandemic.}{In this work we demonstrate how to automate parts of the infectious disease-control policy-making process via performing inference in existing epidemiological models.  The kind of inference tasks undertaken include computing the posterior distribution over controllable, via direct policy-making choices, simulation model parameters that give rise to acceptable disease progression outcomes.  Among other things, we illustrate the use of a probabilistic programming language that automates inference in existing simulators.  Neither the full capabilities of this tool for automating inference nor its utility for planning is widely disseminated at the current time.  Timely gains in understanding about how such simulation-based models and inference automation tools applied in support of policy-making could lead to less economically damaging policy prescriptions, particularly during the current COVID-19 pandemic.} 
\end{abstract}


\section{Introduction}
\label{sec:introduction}
Our goal in this paper is to demonstrate how the \textit{``planning as inference''} methodology
\changed{and}{at the intersection of Bayesian statistics and optimal control}
can directly aid policy-makers in assessing policy options and achieving policy goals\changed{}{, when implemented using epidemiological simulators and suitable automated software tools for probabilistic inference}.  Such software tools can be used to quickly identify the range of values \changed{}{towards which} controllable variables should be driven by \changed{law}{means of policy interventions}, social pressure, or public messaging, so as to limit the spread and impact of an infectious disease such as COVID-19.  

In this work, we introduce and apply a simple form of planning as inference in epidemiological models to automatically identify policy decisions that achieve a desired, high-level outcome.  As but one example, if our policy aim is to \changed{produce}{contain} infectious population totals \changed{that remain}{} below some threshold at all times in the \changed{}{foreseeable} future, we can condition on this putative future holding and examine the allowable \changed{posterior distribution}{values} of controllable behavioural variables at the agent or population level, \changed{}{which in the framework of planning as inference is formalized in terms of a \textit{posterior distribution}}.  As we already know, to control the spread of COVID-19 and its impact on society, policies must be enacted that reduce disease transmission probability or lower the frequency and size of social interactions. This is because we might like to, for instance, not have the number of infected persons requiring hospitalization exceed the number of available hospital beds.

\changed[R1.Q2, R3.Q2]{}{Throughout this work, we take a Bayesian approach, or at the very least, a \textit{probabilistic} consideration of the task.  Especially in the early stages of a new outbreak, the infectious dynamics are not known precisely.  Furthermore, the spread of the disease cannot be treated deterministically, as a result of either fundamental variability in the social dynamics that drive infections, or of uncertainty over the current infection levels, or of uncertainty over appropriate models for analyzing the dynamics.  Therefore, developing methods capable of handling such uncertainty correctly will allow for courses of action to be evaluated and compared more effectively, and could lead to ``better'' policies: a policy that surpasses the desired objective with $55\%$ probability, but fails with $45\%$ probability, may be considered ``worse'' than a policy that simply meets the objective with $90\%$ probability.  Bayesian analysis also offers a form of probabilistic inference in which the contribution of individual variables to the overall uncertainty can be identified and quantified.  Beyond simply obtaining ``the most effective policy measure,'' this may be of interest to analysts trying to further understand \textit{why} certain measures are more effective than others.}

\changed[R3]{}{We first show how the problem of policy planning can be formulated as a Bayesian inference task in epidemiological models.  This framing is general and extensible.  We then demonstrate how particular existing software tools can be employed to perform this inference task in preexisting stochastic epidemiological models, without modifying the model itself or placing restrictions on the models that can be analyzed.  This approach is particularly appealing, as it decouples the specification of epidemiological models by domain experts from the computational task of performing inference.  This shift allows for more expressive and interpretable models to be expediently analyzed, and the sophistication of inference algorithms to be adjusted flexibly.  }

As a result, the techniques and tools we review in this paper are applicable to simulators ranging from simple \changed{population-scale simulators}{compartmental models} to highly expressive agent-based population dynamics models.  In the former, the controls available to policy-makers are blunt -- e.g., ``reduce social interactions by some fractional amount'' -- but how best to achieve this is left as an exercise for policy-makers.  In the latter, variables like ``probability of individuals adhering to self-isolation'' and ``how long should schools be closed if at all'' can be considered and evaluated in combination and comparison to others as potential fine-grained controls that could achieve the same policy objective more efficiently.  

When governments impose any such controls, both citizens and financial markets want to know how draconian these measures must be and for how long they have to be in effect.  Policy analysis based on models that reflect variability in resources such as healthcare facilities in different jurisdictions could hopefully make the answers to these questions more precise and the controls imposed more optimal in the utility maximizing sense.  The same holds for the difference between models that can or cannot reflect variations in population mixing between rural and urban geographic areas.  A person living in a farming county in central Illinois might reasonably wonder if it is as necessary to shelter in place there as it is in Chicago. 

Current approaches to model-based policy-making are likely to be blunt. \changed{Simple models, for example compartmental models, are rapid to fit to new diseases and easy to compute, but lack the necessary expressivity to consider and evaluate fine-grained policy options, for instance regional policy-making.}{Simple models, e.g., compartmental models with few compartments, are rapid to fit to new diseases and easy to compute, but are incapable of evaluating policy options that are more fine-grained than the binning used, such as regionalized measures.}
The net effect of being able to only consider blunt controls arguably has \changed{led}{contributed} to a collective dragging of feet, even in the face of the current COVID-19 pandemic. This delayed reaction, combined with brute application of control, has led to devastating socioeconomic impact, with many sectors such as education, investment markets and small-medium enterprises being directly impacted.

We can and should be able to do better. We believe, and hope to demonstrate, that models and software automation focused on planning as inference for policy analysis and recommendation is one path forward that can help us better react to this and future pandemics, and improve our public health preparedness. 

We upfront note that the specific models that we use in this paper are far from perfect. First, the pre-existing models we use to demonstrate our points in this paper are \changed{not perfectly}{only crudely} calibrated to present-day population dynamics and specific COVID-19 characteristics. We have made some efforts on the latter point, in particular sourcing a COVID-19 adapted compartmental model \cite{kermack1927contribution, blackwood2018introduction, sei3r2020website} and parameters from \cite{FER-20, MAG-20, RIO-20-China, TRA-20-Italy, RUS-20-Italy, PUJ-20-India, PEN-20-China,MAS-20-France,ROV-20-italy, WEN-20-China}, but we stress this limitation.  In addition, \changed{}{in simple cases}, the type of problems for which we discuss solutions in this paper may be solved with more \changed{direct}{straightforward} implementations involving parameter sweeps and ``manual'' checking for desired policy effects, albeit at potentially higher human cost. \changed{}{In this sense, our goal is not to claim fundamental novelty, uniqueness or superiority of any particular inference technique, but rather to raise awareness for the practical feasibility of the Bayesian formulation of planning as inference, which offers a higher level of flexibility and automation than appears to be understood widely in the policy-making arena.}

Note also that current automated inference techniques for stochastic simulation-based models \citep{kypraios2017tutorial,mckinley2018approximate,toni2009approximate,minter2019approximate,chatzilena2019contemporary} are computationally demanding and are by their very nature approximate.
The academic topic at the core of this paper and the subject of a significant fraction of our academic work \citep{NAD-19,ZHO-19,le2016inference,paige2016inference,rainforth2017nesting,rainforth2016interacting,Tolpin-ECMLPKDD-2015, vandeMeent-AISTATS-2015, Paige-NIPS-2014,Wood-AISTATS-2014, WAR-20} deals
with this challenge. Furthermore, the basic structure of simulators currently available may lack important policy-influenceable interaction parameters that one might like to examine. If viewed solely in light of the provided examples, our contribution could reasonably be seen both as highlighting the utility of inference for planning in this application setting, and as automating the manual selection of \changed{}{promising} policy parameters.  The tools we showcase are capable of significantly more; however, for expediency and clarity, we have focused on control as inference, an application that has seen relatively little specific coverage in the literature,
\changed{}{and the simplest possible inference methods which do not require familiarity with the technical literature on approximate Bayesian inference}.  We leave other straightforward applications of automated inference tools in this application area, like parameter inference from observed outbreak data \citep{kypraios2017tutorial,mckinley2018approximate,toni2009approximate,minter2019approximate,chatzilena2019contemporary}, to others. 

That being said, our hope is to inform field epidemiologists and policy-makers about an existing technology that could, right now, be used to support public policy planning towards more precise, potentially tailored interventions that ensure safety while also potentially leading to fewer economic ramifications.  Fully probabilistic methods are apparently only relatively recently being embraced by the epidemiology community \citep{funk2020choices,lessler2016trends}, while the communities for approximate Bayesian inference and simulation-based inference have remained mostly focused on the tasks of parameter estimation and forecasting  \citep{kypraios2017tutorial,mckinley2018approximate,toni2009approximate}, rather than control as inference. Beyond this demonstration, we hope to encourage timely and significant developments on the modelling side, and, if requested, to actually aid in the fight against COVID-19 by helping arm policy-makers with a new kind of tool and training them how to use it rapidly.  Finally, we hope to engage the machine learning community in the fight against COVID-19 by helping translate between the specific epidemiological control problem and the more general control problem formulations on which we work regularly.

\section{Assumptions and Findings}
\label{sec:assumptions}

We start with the assumption that \changed{}{the effectiveness of} policy-making can be \changed{most effective when informed by}{significantly improved by consulting the} outputs of model-based tools which \changed{quantitatively examine}{provide quantitative metrics for} the ability of particular policy actions to achieve specific \changed{pre-defined}{formalized} goals.
In particular, we imagine the following scenario. There exists some current population, and the \changed{}{health} status of its constituents \changed{with respect to disease status}{} is only partially known. There exists a disease whose transmission properties may \changed{}{be} only partially \changed{be}{} known, but whose properties cannot themselves be readily controlled. There exists a population dynamic that can \changed{, in part,}{} be controlled \changed{}{in some limited ways at the aggregate level}. There exists a ``policy goal'' or target which we will refer to as the allowable, allowed, or goal set \changed{}{of system trajectories}. An example of this could be ``the total number of infectious persons should not ever exceed some percentage of the population'' or ``the first date at which the total number of infectious persons exceeds some threshold is at least some number of days away.'' We \changed{}{finally} assume an implied \changed{}{set of} allowable \changed{population dynamics}{policy prescriptions}, \changed{}{defined} in the sense that \changed{if a}{} population \changed{behaves}{dynamics behaving} according to \changed{a prescribed policy, the goal}{such policies} will be \changed{attained}{exactly the ones to attain the goal} with high probability. \changed{and if the population does not follow the policy, the goal will not be attained with high probability.}{}  \changed{}{In general, this set of allowable policies is intractable to compute exactly, motivating the use of automated tools implementing well established \textit{approximate Bayesian inference} methods.} \changed{We acknowledge but ignore the costs and benefits of choosing among policies and suggest that many approaches to maximizing utility require solving the problem we address in this paper as a sub-problem first.}{We explicitly do not claim ``completeness'' of the stochastic dynamic models in any realistic sense, disregarding complications such as potential agent behaviour in strategic response to regionalized policies, and do not attempt to quantify all costs and benefits of the considered policies, for example economic or cultural impacts. Rather, formulating the problem described above in terms of Bayesian inference results in a \textit{posterior distribution} over policies which have been conditioned on satisfying the formalized policy desiderata within the formalized dynamical model. Effectively, this is to be understood as ``scrutinizing,'' ``weighting,'' ``prioritizing,'' or ``focusing'' potential policy actions for further consideration, rather than as an ``optimal'' prescription. When selecting a policy based on the posterior distribution, policy-makers are expected to account for additional, more complex socio-economic phenomena, costs and benefits using their own judgment.}

The \textit{only} things that may safely be taken away from this paper are the following:

\begin{itemize}
    \item Existing \changed{}{compartmental models and} agent-based \changed{and other}{} simulators can be used \changed{}{as an aid} for \changed{planning by framing the planning problem as inference}{policy assessment via a Bayesian planning as inference formulation}.
    \item \changed{}{Existing} automated inference tools can be used to perform the required \changed{inference}{inferential computation}.
    \item Opportunities exist for various fields to come together to improve both understanding of and availability of these techniques and tools.
    \item Further research and development into modelling and inference is recommended to be immediately undertaken to explore the possibility of more efficient, less economically devastating control of the COVID-19 pandemic.  
\end{itemize}

What should {\textit not} be taken away from this paper are any other conclusions, including in particular the following:

\begin{itemize}
    \item Any conclusion or statements that there might exist less aggressive measures that could still be effective in controlling \cov{}.
    \item \changed{}{Any substantial novelty, uniqueness or performance claims about the particular numerical methods and software implementations for Bayesian inference which were used for the purpose of demonstrating the findings above. In particular, on the one hand, similar results could have been obtained using other software implementation strategies in principle, and on the other hand, more advanced inference methods could have been applied using the same software tools at the expense of rendering the conceptual exposition less accessible to audiences outside of the Bayesian inference community.}
\end{itemize}

\changed{}{As scientists attempting to contribute ``across the aisle,''}
we use more qualifying statements than usual throughout this work in an attempt to
\changed{avoid potentially inappropriate headlines. As scientists trying to contribute ``across the aisle,'' we are simply trying to avoid}{reduce the risk of misunderstandings and sensationalism}.

\section{Approach}
\label{sec:approach}

In this section we formalize the policy-making task in terms of computing conditional probabilities in a probabilistic model of the disease spread. While the technical description can get involved at times, we emphasize that in practice the probabilistic model is already defined by an existing epidemiological simulator and the probabilistic programming tools we describe in this paper provide the ability to compute the required conditional probabilities automatically, including automatically introducing required approximations, so the users only need to focus on identifying which variables to condition on, and feeding real-world data to the system.  Readers familiar with framing planning as inference may wish to skip directly to Section~\ref{sec:models}.

Being able to perform probabilistic inference is crucial for taking full advantage of available simulators in order to design good policies. This is because in the real world, many of the variables crucially impacting the dynamics of simulations are not directly observable, such as the number of infectious but asymptomatic carriers or the actual number of contacts people make in different regions each day. These variables are called \emph{latent}, as opposed to \emph{observable} variables such as the number of deaths or the number of passengers boarding a plane each day, which can often be directly measured with high accuracy. It is often absolutely crucial to perform inference over some latent variables to achieve reliable forecasts. For example, the future course of an epidemic like COVID-19 is driven by the number of people currently infected, rather than the number of people currently hospitalized, while in many countries in the world currently only the latter is known.

While performing inference over latent variables is very broadly applicable, the scenario described above being but one example, in this paper we primarily address the problem of choosing good policies to reduce the impact of an epidemic which can also be formulated as an inference problem.  This choice of problem was driven by the hypothesis that the search for effective controls may not in fact be particularly well-served by automation at the current time.  In the epidemiological context, the questions we are trying to answer are ones like ``when and for how long do we need to close schools to ensure that we have enough ventilators for everyone who needs them?'' While obviously this is overly simplistic and many different policy decisions need to be enacted in tandem to achieve good outcomes, we use this example to illustrate tools and techniques that can be applied to problems of realistic complexity. 

Our approach is not novel, it has been studied extensively under the name ``control via planning as inference'' and is now well understood \citep{todorov2008general,toussaint2009robot,kappen2012optimal,levine2018reinforcement}. What is more, the actual computations that result from following the recipes for planning as inference can be, in some cases readily, manually replicated. Again, our aim here is to inform or remind a critically important set of policy-makers and modellers that these methodologies are extremely relevant to the current crisis. Moreover, at least partial automation of model-informed policy-guidance is achievable using existing tools, and, may even lead to sufficient computational savings to make their use in current policy-making practical. Again, our broader hope here is to encourage rapid collaborations leading to more targeted and less-economically-devastating policy recommendations.

\subsection{An Abstract Epidemiological Dynamics Model}
\label{sec:approach:abstract-model}

In this work we will look at both compartmental and agent-based models.  An overview of these specific types of models appears later. For the purposes of understanding our approach to planning as inference, it is helpful to describe the planning as inference problem in a formalism that can express both types of models. The approach of conducting control via planning as inference follows a general recipe:

\begin{enumerate}
    \item Define the latent and control parameters of the model and place a prior distribution over them.
    \item Either or both define a likelihood for the observed disease dynamics data and design constraints that define acceptable disease progression outcomes.
    \item Do inference to generate a posterior distribution on control values that conditions both on the observed data and the defined constraints.
    \item Make a policy recommendation by picking from the posterior distribution consisting of effective control values according to some utility maximizing objective.
\end{enumerate}
 
 We focus on steps 1-3 of this recipe, and in particular do not explore simultaneous conditioning. We ignore the observed disease dynamics data and focus entirely on inference with future constraints.  We explain the rationale behind these choices near the end of the paper.

Very generally, an epidemiological model consists of a set of global parameters and time dependent variables. Global parameters are $(\theta, \eta)$, where $\theta$ denotes parameters that can be controlled by policy directives (e.g.~close schools for some period of time or decrease the general level of social interactions by some amount), and $\eta$ denotes parameters which can not be affected by such measures (e.g.~the incubation period or fatality rate of the disease). 

The time dependent variables are $(X_t, Y_t, Z_t)$ and jointly they constitute the full state of the simulator. $X_t$ are the latent variables we are doing inference over (e.g.~the total number of infected people or the spatio-temporal locations of outbreaks), $Y_t$ are the observed variables whose values we obtain by measurements in the real world (e.g.~the total number of deaths or diagnosed cases), and $Z_t$ are all the other latent variables whose values we are not interested in knowing (e.g.~the number of contacts between people or hygiene actions of individuals). For simplicity, we assume that all variables are either observed at all times or never, but this can be relaxed.

The time $t$ can be either discrete or continuous. In the discrete case, we assume the following factorization
\begin{align}
    p(\theta, \eta, X_{0:T}, Y_{0:T}, Z_{0:T}) =
    p(\theta) p(\eta) p(X_0, Y_0, Z_0 \mid \theta, \eta)
    \prod_{t=1}^T p(X_t, Y_t, Z_t \mid X_{0:t-1}, Y_{0:t-1}, Z_{0:t-1}, \theta, \eta) .
\end{align}

Note that we do not assume access to any particular factorization between observed and latent variables. We assume that a priori the controllable parameters $\theta$ are independent of non-controllable parameters $\eta$ \changed{to ensure that conditioning on desired properties of the epidemic performed in Section \ref{sec:control-inference} has the intended effect of affecting controllable parameters only}{to avoid situations where milder control measures are associated with better outcomes because they tend to be deployed when the circumstances are less severe, which would lead to erroneous conclusions when conditioning on good outcomes in Section \ref{sec:control-inference}}.

\subsection{Inference} \label{sec:inference}

The classical inference task \citep{kypraios2017tutorial,mckinley2018approximate,toni2009approximate,minter2019approximate,chatzilena2019contemporary} is to compute the following conditional probability 
\begin{align}
    p(\eta, X_{0:T} \mid Y_{0:T}, \theta) =
    \int p(\eta, X_{0:T}, Z_{0:T} \mid Y_{0:T}, \theta) \ dZ_{0:T}  .
\end{align}
In the example given earlier $X_t$ would be the number of infected people at time $t$ and $Y_t$ would be the number of hospitalized people at time $t$. If the non-controllable parameters $\eta$ are known they can be plugged into the simulator, otherwise we can also perform inference over them, like in the equation above. This procedure automatically takes into account prior information, in the form of a model, and available data, in the form of observations. It produces estimates with appropriate amount of uncertainty depending on how much confidence can be obtained from the information available.

The difficulty lies in computing this conditional probability, since the simulator does not provide a mechanism to sample from it directly and for all but the simplest models the integral cannot be computed analytically. The main purpose of probabilistic programming tools is to provide a mechanism to perform the necessary computation automatically, freeing the user from having to come up with and implement a suitable algorithm. In this case, approximate Bayasian computation (ABC) would be a suitable tool. We describe it below, emphasizing again that its implementations are already provided by existing tools \citep{kypraios2017tutorial,mckinley2018approximate,toni2009approximate,minter2019approximate,chatzilena2019contemporary}.

The main problem in this model is that we do not have access to the likelihood $p(Y_t \mid X_t, \theta, \eta)$ so we can not apply the standard importance sampling methods. To use ABC, we extend the model with auxiliary variables $Y^{obs}_{0:T}$, which represent the actual observations recorded, and use a suitably chosen synthetic likelihood $p(Y^{obs}_t | Y_t)$, often Gaussian. Effectively, that means we're solving the following inference problem,
\begin{align}
    p(X_{0:T} \mid Y^{obs}_{0:T}, \theta) =
    \iiint p(\eta, X_{0:T}, Y_{0:T}, Z_{0:T} \mid Y^{obs}_{0:T}, \theta) \ dY_{0:T} \ dZ_{0:T} \ d\eta ,
\end{align}
which we can solve by importance sampling from the prior. Algorithmically, this means independently sampling a large number $N$ of trajectories from the simulator
\begin{align}
    (\eta^{(i)}, X^{(i)}_{0:T}, Y^{(i)}_{0:T}, Z^{(i)}_{0:T})
    \stackrel{\text{i.i.d.}}{\sim}
    p(\eta, X_{0:T}, Y_{0:T}, Z_{0:T} \mid \theta) \quad \text{for} \ i \in \{1,\dots,N\} ,
\end{align}
computing their importance weights
\begin{align}
    w_i = \frac{p(Y^{obs}_{0:T} \mid Y^{(i)}_{0:T})}{\sum_{j=1}^N p(Y^{obs}_{0:T} \mid Y^{(j)}_{0:T})} ,
\end{align}
and approximating the posterior distribution
\begin{align}
    p(X_{0:T} \mid Y_{0:T}, \theta) \approx \hat{p}(X_{0:T} \mid Y_{0:T}, \theta) = \sum_{i=1}^N w_i \delta_{X^{(i)}_{0:T}}(X_{0:T}) ,
\end{align}
where $\delta$ is the Dirac delta putting all the probability mass on the point in its subscript. In more intuitive terms, we are approximating the posterior distribution with a collection of weighted samples where weights indicate their relative probabilities.

\subsection{Control as Inference: Finding Actions That Achieve Desired Outcomes} \label{sec:control-inference}

In traditional inference tasks we condition on data observed in the real world. In order to do control as inference, we instead condition on what we \emph{want} to observe in the real world, which tells us which actions are likely to lead to such observations. This is accomplished by introducing auxiliary variables that indicate how desirable a future state is or is not. In order to keep things simple, here we restrict ourselves to the binary case where $Y_t \in \{0,1\}$, where $1$ means that the situation at time $t$ is acceptable and $0$ means it is not. This indicates which outcomes are acceptable, allowing us to compute a distribution over those policies, while leaving the choice of which specific policy likely to produce an acceptable outcome to policymakers. For example, $Y_t$ can be $1$ when the number of patients needing hospitalization at a given time $t$ is smaller than the number of hospital beds available and $0$ otherwise.

To find a policy $\theta$ that is likely to lead to acceptable outcomes, we need to compute the posterior distribution
\begin{align}
    p\left(\theta \mid \forall_t: Y_t = 1\right) \label{eq:control-inference} ,
\end{align}
Once again, probabilistic programming tools provide the functionality to compute this posterior automatically. 
\changed[]{
In this case, rejection sampling would be an appropriate algorithm, which repeatedly samples values of $\theta$ from the prior $p(\theta)$, runs the full simulator using $\theta$, and keeps the sampled $\theta$ only if all $Y_t$ are $1$. The collection of accepted samples approximates the desired posterior.
}{
In this case, rejection sampling would be a simple and appropriate inference algorithm. The rejection sampling algorithm repeatedly samples values of $\theta$ from the prior $p(\theta)$, runs the full simulator using $\theta$, and keeps the sampled $\theta$ only if all $Y_t$ are $1$. The collection of accepted samples approximates the desired posterior. We use rejection sampling in our agent-based modeling experiments, but emphasize that other, more complex and potentially more computationally efficient, approaches to computing this posterior exist.
}

This tells us which policies are most likely to lead to a desired outcome but not how likely a given policy is to lead to that outcome. To do that, we can evaluate the conditional probability $p(\forall_t: Y_t = 1 \mid \theta)$, which is known as the model evidence, for a particular $\theta$. A more sophisticated approach would be to condition on the policy leading to a desired outcome with a given probability $p_0$, that is
\begin{align}
    p\left(\theta \mid p\left(\forall_t: Y_t = 1 \mid \theta\right) > p_0\right) \label{eq:nested} .
\end{align}
For example, we could set $p_0 = 0.95$ to find a policy that ensures availability of hospital beds for everyone who needs one with at least $95\%$ probability. The conditional probability in Equation \ref{eq:nested} is more difficult to compute than the one in Equation \ref{eq:control-inference}. It can be approximated by methods such as nested Monte Carlo (NMC) \citep{rainforth2017nesting}, which are natively available in advanced probabilistic programming systems such as Anglican \citep{tolpin2016design} and WebPPL \citep{webppl} but in specific cases can also be implemented on top of other systems, such as PyProb \citep{le-2016-inference}, with relatively little effort, although using NMC usually has enormous computational cost.

To perform rejection sampling with nested Monte Carlo, we first draw a sample $\theta_i \sim p(\theta)$, then draw $N$ samples of $Y^{(j)}_{0:T} \stackrel{\text{i.i.d.}}{\sim} p(Y_{0:T} \mid \theta_i)$ and reject $\theta_i$ if fewer than $p_0N$ of sampled sequences of $Y$s are all $1$s, otherwise we accept it. This procedure is continued until we have a required number $K$ of accepted $\theta$s. For sufficiently high values of $N$ and $K$, this algorithm approximates the posterior distribution \eqref{eq:nested} arbitrarily well.

However we compute the posterior distribution, it contains multiple values of $\theta$ that represent different policies that, if implemented, can achieve the desired result. In this setup it is up to the policymakers to choose a policy $\theta^*$ that has support under the posterior, i.e.~yields the desired outcomes, taking into account some notion of utility. 

\changed[R3.Q3 - explaning scalability of inference vs grid search]{}{Crucially, despite their relative simplicity, the rejection sampling algorithms we have discussed evaluate randomly sampled values of $\theta$. This is a fundamental difference from the commonly used grid search over a deterministic array of parameter values. In practical terms, this is important because “well distributed” random samples are sufficient for experts to gauge the quantities of interest, and avoid grid searches that would be prohibitively expensive for $\theta$ with more than a few dimensions.}

\subsection{\changed{}{Stochastic} Model Predictive Control: Reacting to What's Happened} 
\label{sec:mpc}

During an outbreak governments continuously monitor and assess the situation, adjusting their policies based on newly available data.  A convenient theoretical and general framework to formalize this is that of model predictive control~\citep{camacho2013model}.  In this case, $Y_t$ consists of variables $Y^{data}_t$ that can be measured as the epidemic unfolds (such as the number of deaths) and the auxiliary variables $Y^{aux}_t$ that indicate whether desired goals were achieved, just like in Section \ref{sec:control-inference}. Say that at time $t=0$ the policymakers choose a policy to enact $\theta^*_0$ based on the posterior distribution
\begin{align}
    p(\theta \mid \forall_{t > 0} : Y^{aux}_t = 1) = \iint p(\theta \mid X_0, Y^{data}_0, Z_0, \forall_{t > 0} : Y^{aux}_t = 1) p(X_0, Z_0) \ dX_0 \ dZ_0  .
\end{align}
Then at time $t=1$ they will have gained additional information $Y^{data}_1$, leading to a new belief over the current state that we denote as $\hat{p}_1(X_1,Z_1)$, for which we give a formula in the general case in Equation \ref{eq:belief-state}. The policymakers then choose the policy $\theta^*_1$  from the posterior distribution $p(\theta \mid \forall_{t > 1} : Y^{aux}_t = 1)$. 

Generally, at time $t$ we compute the posterior distribution conditioned on the current state and achieving desirable outcomes in the future
\begin{align}
    p(\theta \mid \forall_{t' > t} : Y^{aux}_{t'} = 1) = \iint p(\theta \mid X_{t}, Y^{data}_{t}, Z_{t}, \forall_{t' > t} : Y^{aux}_{t'} = 1) \hat{p}_{t}(X_{t}, Z_{t}) \ dX_{t} \ dZ_{t}  . \label{eq:mpc}
\end{align}
Policymakers then can use this distribution to choose the policy $\theta^*_{t}$ that is actually enacted.
The current belief state is computed by inference
\begin{align}
    \hat{p}(X_t, Z_t) = \iint p(X_t, Z_t \mid Y_t, \theta^*_t, \eta, X_{t-1}, Y_{t-1}, Z_{t-1}) \hat{p}(X_{t-1}, Z_{t-1}) \ dX_{t-1} \ dZ_{t-1} . \label{eq:belief-state} 
\end{align}
Equation \ref{eq:belief-state} can be computed using methods described in Section \ref{sec:inference}, while Equation \ref{eq:mpc} can be computed using methods described in Section \ref{sec:control-inference}.  As such, the policy enacted evolves over time, as a result of re-solving for the optimal control based on new information.  

\changed[R1.Q1.minor.1]{}{We note that what we introduce here as MPC may appear to be slightly different from what is commonly referred to as \emph{model predictive control}~\citep{garcia1989model}.  Firstly, instead of solving a finite-dimensional optimization problem over controls at each step, we perform a Bayesian update step and sample from the resulting posterior distribution over controls.
Secondly, in more traditional applications of MPC a \emph{receding horizon}~\citep{mattingley2011receding} is considered, where a finite and fixed-length window is considered.  The controls required to satisfy the constraint over that horizon are then solved for and applied -- without consideration of timesteps beyond this fixed window.  At the next time step, the controls are then re-solved for.  In this work, we rather consider a constant policy for the remaining $T-t$ time steps, as opposed to allowing a variable policy (discussed below) over a fixed window.  We note that time-varying controls are fully permissible under the framework we present, as we demonstrate in Section \ref{sec:approach:time_varying_control}.  Furthermore, we could easily consider a fixed horizon, just by changing the definition of $Y_{t'}^{aux}$ to be equal to one for $t' > t + p$, where $p$ is the length of the window being  considered.  Both of these extensions are provisioned for under the framework we provide, and could be used to implement what might be considered as more conventional stochastic MPC, but with all the auxiliary benefits of Bayesian inference.  }  

\changed[R1 \& R3]{}{Before we proceed, it is critical to note that the posterior $p(\theta \mid \forall_{t > 1} : Y^{aux}_t = 1)$ describes the probability, given the dynamic model and the currently available information, for the parameters $\theta$ to achieve the policy goals.  As such, our approach provides a method for ``screening'' policies such that candidate policies that achieve the required outcomes are found.  What it does not take into account is the relative ``cost'' of each policy, or, how best to achieve the required parameter values.  For instance, in the \fancyseir{} examples we present later, we learn the reduction in transmission required to avoid exceeding an infection threshold.  The reduction in transmission can be achieved (for instance) by encouraging hand-washing and social distancing.  While the method will identify whether or not a particular level of transmission will achieve the desired outcome, it does not indicate the optimal trade-off between the cost of a particular policy, the amount that the policy exceeds the required outcome, and the probability of the outcome.  A human must therefore select, from the set of compliant policies identified, the policy that maximizes utility while minimizing cost.  Therefore, this method is an important component of a wider policy making toolkit, as opposed to an oracle that can dictate the optimal policy decisions.  }

\subsection{Time-Varying Control: Long Term Planning}\label{sec:approach:time_varying_control}

It is also possible to explicitly model changing policy decisions over time, which enables more detailed planning, such as when to enact certain preventive measures such as closing schools. Notationally, this means instead of a single $\theta$ there is a separate $\theta_t$ for each time $t$. We can then find a good sequence of policy decisions by performing inference just like in Section \ref{sec:control-inference} by conditioning on achieving the desired outcome
\begin{align}
\label{eq:condition-outcome}
    p(\theta_{0:T} \mid \forall{t}: \, Y_t = 1) .
\end{align}
The inference problem is now more challenging, since the number of possible control sequences grows exponentially with the time horizon. Still, the posterior can be efficiently approximated with methods such as Sequential Monte Carlo.

It is straightforward to combine this extension with model predictive control from Section \ref{sec:mpc}. The only required modification is that in Equation \ref{eq:mpc} we need to condition on previously enacted policies and compute the posterior over all future policies.
\begin{align}
    p(\theta_{t+1:T} \mid \forall_{t' > t} \ Y^{aux}_{t'} = 1) = \iint p(\theta_{t+1:T} \mid \theta^*_{0:t}, X_{t}, Y^{data}_{t}, Z_{t}, \forall_{t' > t} : Y^{aux}_{t'} = 1) \hat{p}_{t}(X_{t}, Z_{t}) \ dX_{t} \ dZ_{t}  .
\end{align}
At each time $t$ the policymakers only choose the current policy $\theta^*_t$, without committing to any future choices. This combination allows for continuously reevaluating the situation based on available data, while explicitly planning for enacting certain policies in the future.  While here we explicitly consider only open-loop control policies, this re-planning allows new information to be taken into account, and facilitates reactionary policy decisions to evolving situations.  

In models with per-timestep control variables $\theta_t$, it is very important that in the model (but not in the real world) the enacted policies must not depend on anything else. If the model includes feedback loops for changing policies based on the evolution of the outbreak, it introduces positive correlations between lax policies and low infection rates (or other measures of severity of the epidemic), which in turn means that conditioning on low infection rates is more likely to produce lax policies. This is a known phenomenon of reversing causality when correlated observational data is used to learn how to perform interventions \cite{pearl-causality}.

\changed[R1.Q5.4]{}{We note another potential pitfall that may be exacerbated by time-varying control: if the model used does not accurately reflect reality, any form of model-based control is likely to lead to poor results. To some extent, this can be accounted for with appropriate distributions reflecting uncertainty in parameter values. However, given the difficulty of modeling human behavior, and especially of modeling people's reactions to novel policies, there is likely to be some mismatch between modeled and real behavior. To give an example of a possible flaw in an agent-based model: if a proposed fine-grained policy closed one park while keeping open a second, nearby, park, the model may not account for the likely increase in visitors to the second park. The larger and more fine-grained (in terms of time or location), the space of considered policies is, the more such deficiencies may exist. We therefore recommend that practitioners restrict the space of considered policies to those which are likely to be reasonably well modeled.}

\subsection{Automation}

We have intentionally not really explained how one might actually computationally characterize any of the conditional distributions defined in the preceding section.  For the compartmental models that follow, we provide code that directly implements the necessary computations.  Alternatively we could have used the automated inference facilities provided by any number of packages or probabilistic programming systems.  Performing inference as described in existing, complex simulators is much more complex and not nearly as easy to implement from scratch.  However, it can now be automated using the tools of probabilistic programming.

\subsubsection{Probabilistic Programming}
\label{sec:probprog}
Probabilistic programming \citep{MEE-18} is a growing subfield of machine learning that aims to build an analogous set of tools for automating inference as automatic differentiation did for continuous optimization.  Like the gradient operator of languages that support automatic differentiation, probabilistic programming languages introduce observe operators that denote conditioning in the probabilistic or Bayesian sense.  In those few languages that natively support nested Monte Carlo \citep{tolpin2016design,dippl}, language constructs for defining conditional probability objects are introduced as well.  Probabilistic programming languages (PPLs) have semantics \citep{staton2016semantics} that can be understood in terms of Bayesian inference \citep{ghahramani2015probabilistic,gelman2013bayesian,bishop2006pattern}. The major challenge in designing useful PPL systems is the development of general-purpose inference algorithms that work for a variety of user-specified programs.  The work in the paper uses only the very simplest, and often least efficient, general purpose inference algorithms, importance sampling and rejection sampling.  Others are covered in detail in \citep{MEE-18}.

Of all the various probabilistic programming systems, only one is readily compatible with inference in existing stochastic simulators: PyProb\citep{LE-19}.  Quoting from its website\footnote{\url{https://github.com/pyprob/pyprob}} ``PyProb is a PyTorch-based library for probabilistic programming and inference compilation. The main focus of PyProb is on coupling existing simulation codebases with probabilistic inference with minimal intervention.''  A textbook, technical description of how PyProb works appears in \citep[Chapt.~6]{MEE-18}.

Recent examples of its use include inference in the standard model of physics conditioning on observed detector outputs \citep{BAY-18,BAY-19}, inference about internal composite material cure processes through a composite material cure simulator conditioned on observed surface temperatures \citep{MUN-20}, and inference about malaria spread in a malaria simulator \citep{gram2019hijacking}.

\subsubsection{\changed[R1.Q5.main.1 (this entire sub-sub-section \ref{sec:alternative_approaches}).]{}{Alternative Approaches}}
\label{sec:alternative_approaches}
It is important to note upfront that our ultimate objective is to understand the dependence of the policy outcome on the controllable parameters (for instance, as defined by Equation \eqref{eq:control-inference} and \eqref{eq:nested}).  There are a litany of methods to quantify this dependency.  Each method imposes different constraints on the model family that can be analyzed, and the nature of the solution obtained.  For instance, the fully Bayesian approach we take aims to quantify the entire distribution, whereas traditional \emph{optimal control} methods may only seek a pointwise maximizer of this probability.  Therefore, before we introduce the models we analyze and examples we use to explore the proposed formulation, we give a brief survey of alternative approaches one could use to solve this problem, and give the benefits and drawbacks of our proposed formulation compared to these approaches.  

The traditional toolkits used to analyze problems of this nature are \emph{optimal control}~\citep{vinter2010optimal, lewis2012optimal, sharomi2017optimal, bertsekas2019reinforcement} and \emph{robust control}~\citep{zhou1998essentials, hansen2001robust, green2012linear}.  Most generally, optimal control solves for the control inputs (here denoted $\theta$) such that some outcome is achieved (here denoted $Y^{aux}_t = 1 \ \forall_{t > 0}$) while minimizing the associated cost of the controls.  An example of this may be successfully maintaining stable flight while minimizing fuel expenditure.  However, for the controls to be optimal, this assumes the model is correct.  Alternatively, robust control does not assume that the model is correct, and instead solves for controls that are maximally robust to the uncertainty in the model, while still achieving the desired outcome. 

Many traditional control approaches can exploit a mathematical model of the system to solve for controls that consider multiple timesteps, referred to as \emph{model-predictive control} (MPC)~\citep{morari1999model, rawlings2000tutorial, camacho2013model}.  Alternatively, control methods can be \emph{model-free}, where there is no notion of the temporal dynamics of the system being controlled, such as the canonical PID control.  Model-based methods often solve for more effective control measures with lower overall computational costs, by exploiting the information encoded in the model.  However, model-predictive control methods are only applicable when a (accurate) model is available.  

One could re-frame the objective as trying to maximize the \emph{reward} of some applied controls.  Here, reward may be a $0$-$1$ indicator corresponding to whether or not the hospital capacity was exceeded at any point in time.  The reward may also be more sophisticated, such as by reflecting the total number of people requiring hospital treatment, weighted by how critical each patient was.  Reward may also internalize some notion of the \emph{cost} of a particular control.  Shutting workplaces may prevent the spread of infection, but has economic and social implications that (at least partially) counteract the benefit of reducing the infection rate.  This can be implemented as earning progressively more negative reward for stronger policies.  

This type of analysis is formalized through \emph{reinforcement learning} (RL)~\citep{sutton1992reinforcement, kaelbling1996reinforcement, sutton2018reinforcement, bertsekas2019reinforcement, warrington2021robust}.  In RL, a \emph{policy}, here $\theta$, is solved for that maximizes the expected reward.  RL is an incredibly powerful toolkit that can learn highly expressive policies that vary as a function of time, state, different objectives etc.  RL methods can also be described as model-based~\cite{moerland2021modelbased} and model-free~\citep{bertsekas2019reinforcement} analogously to traditional control methods.  While we do not delve into RL in detail here, it is important to highlight as an alternative  approach, and refer the reader to \citet{levine2018reinforcement} for a more detailed introduction of RL, \citep{bertsekas2019reinforcement} for discussion of the relationship between RL and control methods, and \citet{levine2018reinforcement} for discussion of the relationship between RL and planning-as-inference.

However, a critical drawback of both control and RL is the strong dependence on the loss functions or specific models considered, both in terms of convergence and the optimal solution recovered~\citep{laud2003influence}.  This dependence may therefore mean that the solution recovered is not representative of the ``true'' optimal solution.  Furthermore, RL approaches can be incredibly expensive to train, especially in high-dimensional time series models, with a large array of controls and sparse rewards.  More traditional control-based methods can be limited in the models that they can analyze, both in terms of the transition distribution and the range of controls that can be considered.  Finally, both RL and control-based approaches offer limited insight into \emph{why} certain controls were proposed.  This can limit the ability of researchers and modellers to expediently investigate the emergent properties of complex, simulation-based models.  Furthermore, this lack of transparency may be disadvantageous for policy-makers, who may be required to justify why certain policy decisions were made, and the confidence with which that policy decision was believed to be optimal.

All of these methods, at least in terms of their core intentions, are very similar. Different methods impose different restrictions on the models that can be analyzed, and the nature of the solution obtained.  In this work, we take the approach of framing control and planning as fully Bayesian inference with a binary objective.  We choose this formulation as it places very few restrictions on the model class that can be analyzed, allows uncertainty to be succinctly handled throughout the inference, a broad range of objectives to be defined under the same framework, and correctly calibrated posterior distributions over all random variables in the model to be recovered.  Furthermore, this formulation allows us to access the ever-growing array of powerful inference algorithms to perform the inference.  While naive approaches such as grid-search or random sampling may be performant in low-dimensional applications, they do not scale to high-dimensions through the \emph{curse of dimensionality}~\citep{koppen2000curse, mcbook}.  Therefore, we focus on methods that can scale to high-dimensional applications~\citep{mcbook}.  Finally, Bayesian inference methods are, arguably, the most complete formulation.  Once the full posterior or joint distribution have been recovered, many different methodologies, constraints, objectives etc can be formulated and evaluated \emph{post facto}.  This flexibility allows for fine-grained analysis to be conducted on the inferred distributions to provide analysts with a powerful and general tool to further analyze and understand the model.  Less myopically, this analysis and the understanding garnered may be more beneficial to our broader understanding of outbreaks and the feasibility and efficacy of particular responses.  

Finally, we note that while throughout the experiments presented in Section \ref{sec:exp:fred} we use the PyProb framework, PyProb is not the \emph{only} framework available.  Any inference methodology could be used to compute the probability in Equation \eqref{eq:control-inference}.  However, PyProb is a natural choice as it allows for greater flexibility in interfacing between black-box simulator code, and powerful and efficiently implemented inference algorithms, without modifying either simulator or inference code.  This reduces the implementation overhead to the analyst, and reduces the scope for implementation bugs and oversights to be introduced.

\section{Models}
\label{sec:models}

Epidemiological dynamics models can be used to describe the spread of a disease such as COVID-19 in society.  Different types  span vastly different levels of fidelity.  There are classical compartmental models (SIR, SEIR, etc.) \citep{JUL-18} that describe the bulk progression of diseases of different fundamental types. These models break the population down to a series of compartments (e.g.~susceptible (S), infectious (I), exposed (E), and recovered (R)), and treat them as as continuous quantities that vary smoothly and deterministically over time following dynamics defined by a particular system of ordinary differential equations. These models are amenable to theoretical analyses and are computationally efficient to forward simulate owing to their low dimensionality. As policy-making tools, they are rather blunt unless the number of compartments is made large enough to reflect different demographic information (age, socio-economic info, etc), spatial strata, or combinations of thereof.

At the other end of the spectrum are agent-based models \cite{HUN-18, hunter2017, TRA-18, BAD-18} (like the Pitt Public Health Dynamics Laboratory's FRED \citep{GRE-13-FRED} or the Institute for Disease Modelling's EMOD \cite{BER-18-EMOD}) that model populations and epidemiological dynamics via discrete agent interactions in ``realistic'' space and time.  Imagine a simulation environment like the game Sim-City$^{TM}$, where the towns, populations, infrastructure (roads, airports, trains, etc.), and interactions (go to work, school, church, etc.) are modelled at a relatively high level of fidelity. These models exist only in the form of stochastic simulators, i.e.~software programs that can be initialized with disease and population characteristics, then run forward to show in a more fine-grained way the spread of the disease over time and space.

Both types of models are useful for policy-making. Compartmental models are usually more blunt unless the number of compartments is very high and it is indexed by spatial location, demographics and age categories. Increasing the number of compartments adds more unknown parameters which must be estimated or marginalized. Agent-based models are complex by nature, but they may be more statistically efficient to estimate, as they are parameterized more efficiently, often directly in terms of actual individual and group behaviour choices. In many cases, predictions made by such models are more high fidelity, certainly more than compartmental models with few compartments, and this has implications for their use as predictive tools for policy analysis. For instance, policies based on simulating a single county in North Dakota with excellent hospital coverage and a highly dispersed, self-sufficient population could lead to different intervention recommendations compared to a compartmental model of the whole of the United States with only a few compartments.

\subsection{A Compartmental Model of COVID-19}
\label{sec:models:seir}

We begin by introducing a low-dimensional compartmental model to explore our methods in a well-known model family, before transitioning to a more complex agent-based simulator. The model we use is an example of a 
classical SEIR model~\cite{kermack1927contribution, blackwood2018introduction, sei3r2020website}. 
In such models, the population is subdivided into a set of compartments, 
representing the susceptible (uninfected), exposed (infected but not yet infectious), infectious (able to infect/expose others) and recovered (unable to be infected). Within each compartment, all individuals are treated identically, and the full state of the simulator is simply the size of the population of each compartment. Our survey of the literature found a lack of consensus about the compartmental model and parameters which most faithfully simulate the COVID-19 scenario. Models used range from standard SEIR \cite{ROV-20-italy, MAS-20-France, TRA-20-Italy, LIU-20}, SIR \cite{PUJ-20, TRA-20-Italy, WEB-20, TEL-20-Portugal, JIA-20}, SIRD \cite{ANA-20, LIU-20b-China, CAC-20}, QSEIR \cite{LIU-20-China}, and SEAIHRD \cite{ARE-20}. The choice depends on many factors, such as how early or late in the stages of an epidemic one is, what type of measures are being simulated, and the availability of real word data. We opted for the model described in this section, which seems to acceptably represent the manifestation of the disease in populations. Existing work has investigated parameter estimation in stochastic SEIR models~\cite{lekone2006stochasticepi, ROBERTS201549}. Although we will discuss how we set the model parameters, we emphasize that our contribution is instead in demonstrating how a calibrated model could be used for planning.

\paragraph{Model description}

\tikzstyle{block} = [rectangle, draw, fill=blue!20, text width=2em, text centered, rounded corners, minimum height=2em]
\tikzstyle{line} = [draw, -latex']
\tikzstyle{cloud} = [draw, ellipse,fill=red!20, node distance=3cm,minimum height=2em]
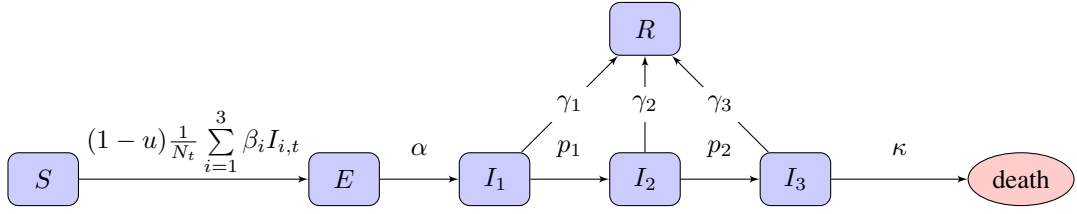
\begin{figure}[t]
    \centering
	\begin{tikzpicture}[every text node part/.style={align=center}, node distance=2cm]
	\node [block, ] (S) {$S$};
	\node [block, right of=S, xshift = 2cm] (E) {$E$};
	\node [block, right of=E] (I1) {$I_1$};
	\node [block, right of=I1] (I2) {$I_2$};
	\node [block, right of=I2] (I3) {$I_3$};
	\node [block, above of=I2] (R) {$R$};
	\node [cloud, right of =I3] (death) {death};
	\path [line] (S) --  node[yshift=.5cm] {$(1-u) \frac{1}{N_t}\sum\limits_{i=1}^3 \beta_i I_{i,t}$} (E);
	\path [line] (E) -- node[yshift=.4cm] {$\alpha$} (I1);
	\path [line] (I1) -- node[yshift=.4cm] {$p_1$} (I2);
	\path [line] (I2) -- node[yshift=.4cm] {$p_2$} (I3);
	\path [line] (I1) -- node[fill=white, yshift=.0cm] {$\gamma_1$} (R);
	\path [line] (I2) -- node[fill=white, yshift=.0cm] {$\gamma_2$} (R);
	\path [line] (I3) -- node[fill=white, yshift=.0cm] {$\gamma_3$} (R);
	\path [line] (I3) -- node[yshift=.4cm] {$\kappa$} (death);
	\end{tikzpicture}
    \caption{Flow chart of the \fancyseir{} model we employ. A member of the susceptible population $S$ 
    moves to exposed $E$ after being exposed to an infectious person, where ``exposure'' is defined as the previous susceptible person contracting the illness. After some incubation period, a 
    random duration parameterized by $\alpha$, they develop a mild infection 
    ($I_1$). They may then either recover, moving to $R$, or progress to a severe infection ($I_2$). From $I_2$, they again may recover, or else progress further to a critical infection ($I_3$). From $I_3$, the critically infected person will either recover or die.}
    \label{fig:SEIR-fig}
\end{figure}

We use an \fancyseir{}~model~\cite{sei3r2020website}, a variation on the standard SEIR model which allows additional modelling freedom. It uses six compartments:  susceptible ($S$), exposed ($E$), infectious with mild ($I_1$), severe ($I_2$) or critical infection ($I_3$), and recovered ($R$). We do not include baseline birth and death rates in the model, although there is a death rate for people in the critically infected compartment. The state of the simulator at time $t \in [0, T]$ is $X_t = \{ S_t, E_t, I_{1,t}, I_{2,t}, I_{3,t}, R_t\}$ with $S_t,$ $E_t,$ $I_{1,t},$ $I_{2,t},$ $I_{3,t},$ and $R_t$ indicating the population sizes (or proportions) at time $t$. The unknown model parameters are $\eta = \{ \alpha, \beta_{1}, \beta_{2}, \beta_{3}, p_{1}, p_{2}, \gamma_{1}, \gamma_{2}, \gamma_{3}, \kappa \}$, each with their own associated prior. To the model we add a free, control parameter, denoted $u\in\left[ 0, 1\right]$, that acts to reduce the transmission of the disease. Since $u$ is the only free parameter, $\theta = u$. An explanation of $u$ is given later in the text. There are no internal latent random variables ($Z_t$) in this model. In this paper we do not demonstrate inference about $\theta$ given $Y^{obs}$ within this model, and so do not consider $Y^{obs}$ here. We do, however, consider $Y^{aux}$ to perform policy selection, and discuss the form of $Y^{aux}$ later. Defining the total live population (i.e. the summed population of all compartments) at time $t$ to be $N_t$, the dynamics are given by the following equations, and shown in Figure~\ref{fig:SEIR-fig}.
\newcommand{\stoe}{(1-u)\frac{1}{N_t}\sum\nolimits_{i=1}^3 \beta_i I_{i,t} S_t}
\newcommand{\etoi}{\alpha E_t}
\newcommand{\itoii}{p_1 I_{1,t}}
\newcommand{\iitoiii}{p_2 I_{2,t}}
\newcommand{\iiitod}{\kappa I_{3,t}}
\newcommand{\itor}{\gamma_1 I_{1,t}}
\newcommand{\iitor}{\gamma_2 I_{2,t}}
\newcommand{\iiitor}{\gamma_3 I_{3,t}}
\begin{minipage}{\linewidth}
\centering
\noindent\begin{minipage}{.45\linewidth}
\begin{alignat}{2}
    &\frac{\mathrm{d}}{\mathrm{d}t}S_t    &&= - \stoe \\
    &\frac{\mathrm{d}}{\mathrm{d}t}E  &&= \stoe - \etoi \\
    &\frac{\mathrm{d}}{\mathrm{d}t}I_{1, t}  &&=  \etoi - \itoii - \itor
\end{alignat}
\end{minipage}
\noindent\begin{minipage}{.45\linewidth}
\begin{alignat}{2}
    &\frac{\mathrm{d}}{\mathrm{d}t}I_{2, t}  &&= \itoii - \iitoiii - \iitor \\
    &\frac{\mathrm{d}}{\mathrm{d}t}I_{3, t}  &&= \iitoiii - \iiitod - \iiitor \\
    &\frac{\mathrm{d}}{\mathrm{d}t}R_t  &&= \itor + \iitor + \iiitor.
\end{alignat}
\end{minipage}
\vspace{.25cm}
\end{minipage}
For the purposes of simulations with this model, we initialize the state with $0.01\%$ of the population having been exposed to the infection, and the remaining $99.99\%$ of the population being susceptible. The population classified as infectious and recovered are zero, i.e. $X_0 = \left\lbrace 0.9999, 0.0001, 0, 0, 0, 0 \right\rbrace$ and $N_t = 1$.

\begin{figure}[t]
    \begin{subfigure}[b]{\textwidth}
    \centering

    \begin{subfigure}[b]{0.48\textwidth}
        \centering
        \includegraphics[width=\textwidth]{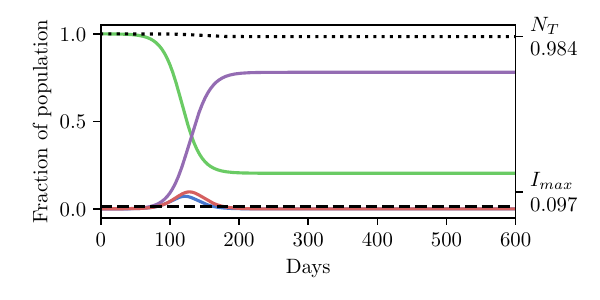}
    \end{subfigure}
    ~
    \begin{subfigure}[b]{0.48\textwidth}
        \centering
        \includegraphics[width=\textwidth]{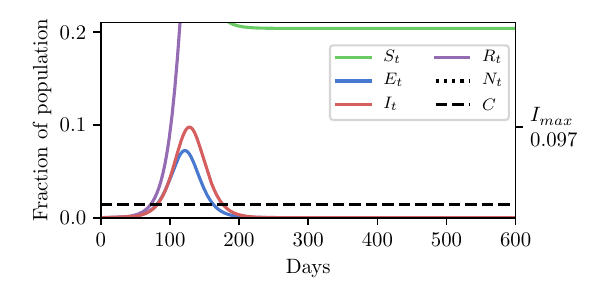}
    \end{subfigure}
    \vspace{-.3cm}
    \caption{ Deterministic trajectory with zero control input ($u = 0$). }
    \label{fig:exp:seir:det_nom}
    \end{subfigure}
    \begin{subfigure}[b]{\textwidth}
    \centering
    \begin{subfigure}[b]{0.48\textwidth}
        \centering        \includegraphics[width=\textwidth]{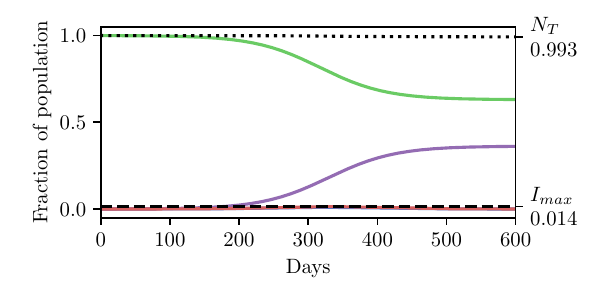}
    \end{subfigure}
    ~
    \begin{subfigure}[b]{0.48\textwidth}
        \centering
        \includegraphics[width=\textwidth]{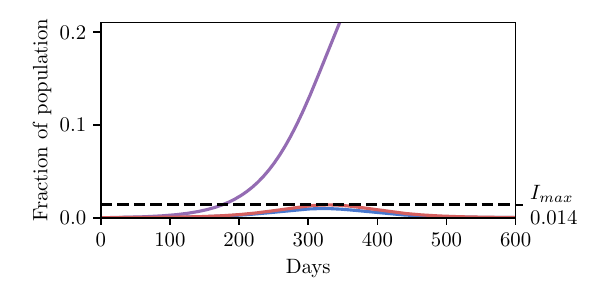}
    \end{subfigure}
    \vspace{-.3cm}
    \caption{ Deterministic trajectory controlled to limit maximum infected population ($u = 0.37$). }
    \label{fig:exp:seir:det_con}
    \end{subfigure}
    \caption{Populations per compartment during deterministic \fancyseir{} simulations, both without intervention (top) and with intervention (bottom). Plots in the left column show the full state trajectory, and in the right column are cropped to more clearly show the exposed and infected populations. Without intervention, the infected population requiring hospitalization ($20\%$ of cases) exceeds the threshold for infected population ($0.0145$, black dashed line), overwhelming hospital capacities. With intervention ($u$=0.37) the infected population always remains below this limit. Note that we re-use the colour scheme from this figure through the rest of the paper.}
    \label{fig:exp_seit:det_nom_con}
\end{figure}

\paragraph{Example trajectories}
Before explaining how we set the \fancyseir{} model parameters, or pose inference problems in the model, we first verify that we are able to simulate feasible state evolutions. As we will describe later, we use parameters that are as reflective of current \cov{} epidemiological data as possible at the time of writing. Figures~\ref{fig:exp:seir:det_nom} and \ref{fig:exp:seir:det_con} show deterministic simulations from the model with differing control values $u$. Shown in green is the susceptible population, in blue is the exposed population, in red is the infectious population, and in purple is the recovered population. The total live population is shown as a black dotted line. All populations are normalized by the initial total population, $N_0$. The dashed black line represents a threshold under which we wish to keep the number of infected people under at all times. The following paragraph provides the rationale for this goal.

\paragraph{Policy goal}
As described in Section~\ref{sec:mpc}, parameters should be selected to ensure that a desired goal is achieved. In all scenarios using the \fancyseir{} model, we aim to maintain the maximal infectious population proportion requiring healthcare below the available number of hospital beds per capita, denoted $C$. This objective can be formulated as an auxiliary observation, $Y_{0:T}^{aux}$, introduced in Section \ref{sec:approach}, as:
\begin{equation}
    Y_{0:T}^{aux} = \mathbb{I}\left[ \left(\max_{t\in 0:T} \left( I_{1,t} + I_{2,t} + I_{3, t} \right)\right) < C\right],
\end{equation}
where $I_{1,0:T}$, $I_{2,0:T}$ and $I_{3,0:T}$ are sampled from the model, conditioned on a $\theta$ value. 
This threshold value we use was selected to be $0.0145$, as there are $0.0029$ hospital beds per capita in the United States~\cite{worldbank-us-beds}, and roughly $20\%$ of \cov{} cases require hospitalization. This constraint was chosen to represent the notion that the healthcare system must have sufficient capacity to care for all those infected who require care, as opposed to just critical patients.  However, this constraint is only intended as a demonstrative example of the nature of constraints and inference questions one can query using models such as these, and under the formalism used here, implementing and comparing inferences under different constraints is very straightforward. 
More complex constraints may account for the number of critical patients differently to those with mild and severe infections, model existing occupancy or seasonal variations in capacity, or, target other metrics such as the number of deceased or the duration of the epidemic.

The constraint is not met in Figure~\ref{fig:exp:seir:det_nom}, but is in Figure~\ref{fig:exp:seir:det_con}, where a greater control input $u$ has been used to slow the spread of the infection. This is an example of the widely discussed ``flattening of the curve.'' As part of this, the infection lasts longer but the death toll is considerably lower. 

\paragraph{Control input}
As noted before, we assume that only a single ``controllable'' parameter affects our model, $u$. This is the reduction in the ``baseline reproductive ratio,'' $R_0$, due to policy interventions. Increasing $u$ has the same effect as reducing the infectiousness parameters $\beta_1$, $\beta_2$ and $\beta_3$ by the same proportion. $u$ can be interpreted as the effectiveness of policy choices to prevent new infections. Various policies could serve to increase $u$, since it is a function of both, for example, reductions in the ``number of contacts while infectious'' (which could be achieved by social distancing and isolation policy prescriptions), and the ``probability of transmission per contact'' (which could be achieved by, e.g., eye, hand, or mouth protective gear policy prescriptions). It is likely that both of these kinds of reductions are necessary to maximally reduce $u$ at the lowest cost.

For completeness, the baseline reproductive ratio, $R_0$, is an estimate of the number of people a single infectious person will in turn infect and can be calculated from other model parameters~\cite{sei3r2020website}. $R_0$ is often reported by studies as a measure of the infectiousness of a disease, however, since $R_0$ can be calculated from other parameters we do not explicitly parameterize the model using $R_0$, but we will use $R_0$ as a convenient notational shorthand. We compactly denote the action of $u$ as controlling the baseline reproductive rate to be a ``controlled reproductive rate,'' denoted $\hat{R}_0$, and calculated as $\hat{R}_0 = (1 - u) R_0$. This is purely for notational compactness and conceptual ease, and is entirely consistent with the model definition above.

\paragraph{Using point estimates of model parameters}
We now explain how we set the model parameters to deterministic estimates of values which roughly match \cov{}. The following section will consider how to include uncertainty in the parameter values. Specifically, the parameters are the incubation period $\alpha^{-1}$; rates of disease progression $p_1$ and $p_2$; rates of recovery from each level of infection, $\gamma_1,$ $\gamma_2,$ and $\gamma_3$; infectiousness for each level of infection, $\beta_1,$ $\beta_2,$ and $\beta_3$; and a death rate for critical infections, $\kappa$. $u \in [ 0, 1 ]$ is a control parameter, representing the strength of action taken to prevent new infections~\cite{boldog2020risk}. To estimate distributions over the uncontrollable model parameters, we consider their relationships with various measurable quantities
\begin{minipage}{\linewidth}
\noindent\begin{minipage}{.4\linewidth}
\begin{alignat}{2}
\label{eq:seir-quantities-begin}
    &\text{incubation period} &&= \alpha^{-1}  \phantom{\frac{1}{1}} \\ 
    &\text{mild duration} &&= \frac{1}{\gamma_1 + p_1} \\
    &\text{severe duration} &&= \frac{1}{\gamma_2 + p_2} \\
    &\text{critical duration}& &= \frac{1}{\gamma_3 + \kappa}
\label{eq:seir-quantities-end-firstcolumn}
\end{alignat}
\end{minipage}
\noindent\begin{minipage}{.6\linewidth}
\begin{alignat}{2}
    &\text{mild fraction} &&= \frac{\gamma_1}{\gamma_1 + p_1} \\
    &\text{severe fraction} &&= \frac{\gamma_2}{\gamma_2 + p_2} \cdot \left( 1 - \text{mild fraction} \right) \\
    &\text{critical fraction} &&= 1 - \text{severe fraction} - \text{mild fraction} \phantom{\frac{1}{1}} \\
    &\text{fatality ratio} &&= \frac{\kappa}{\gamma_3 + \kappa} \cdot (\text{critical fraction}).
\label{eq:seir-quantities-end}
\end{alignat}
\end{minipage}
\vspace{.5cm}
\end{minipage}
Given the values of the left-hand sides of each of Equations \ref{eq:seir-quantities-begin}-\ref{eq:seir-quantities-end}, (as estimated by various studies) we can calculate model parameters $\alpha, p_1, p_2, \gamma_1, \gamma_2, \gamma_3$ and $\kappa$ by inverting this system of Equations. These parameters, along with estimates for $\beta_1$, $\beta_2$, and $\beta_3$, and a control input $u$, fully specify the model. Reference \cite{sei3r2020website} uses such a procedure to deterministically fit parameter values. Given the parameter values, the simulation is entirely deterministic. Therefore, setting parameters in this way enables us to make deterministic simulations of ``typical'' trajectories, as shown in \ref{fig:exp_seit:det_nom_con}. Specifying parameters in this way and running simulations in this system provides a low overhead and easily interpretable environment, and hence is an invaluable tool to the modeller.  

\paragraph{Dealing with uncertainty about model parameter values}

\begin{figure}[t]
    \centering

    \begin{subfigure}[b]{0.48\textwidth}
        \centering
        \includegraphics[width=\textwidth]{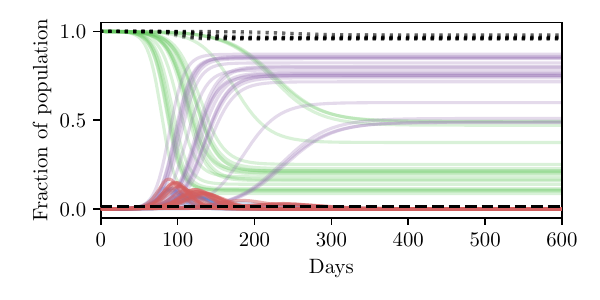}
        \caption{ }
        \label{fig:exp:seir:stoch:full}
    \end{subfigure}
    ~
    \begin{subfigure}[b]{0.48\textwidth}
        \centering
        \includegraphics[width=\textwidth]{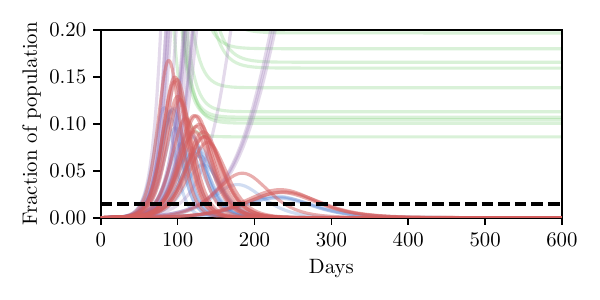}
        \caption{}
        \label{fig:exp:seir:stoch:zoom}
    \end{subfigure}
    
    \caption{Stochastic simulations from the \fancyseir{} model. 
            Figure \ref{fig:exp:seir:stoch:full} shows the full trajectory while Figure \ref{fig:exp:seir:stoch:zoom} is cropped to the pertinent region.  Compared to the deterministic simulations in Figure \ref{fig:exp:seir:det_nom}, stochastic simulations have the capacity to be much more infectious.  \changed[R3.Q3]{}{Therefore, fully stochastic simulations are required to accurately quantify the true risk in light of the uncertainty in the model.  As a result of this, Bayesian methods, or at least methods that correctly handle uncertainty, are required for robust analysis.}  We reuse the colour scheme defined in Figure \ref{fig:exp:seir:det_nom} for trajectories.
            }
    \label{fig:exp:seir:stoch}
\end{figure}

Deterministic simulations are easy to interpret on a high level, but they require strong assumptions as they fix the values of unknown parameters to point estimates. We therefore describe how we can perform inference and conditioning in a stochastic model requiring less strict assumptions, and show that we are able to provide meaningful confidence bounds on our inferences that can be used to inform policy decisions more intelligently than without this stochasticity. As described in Section~\ref{sec:approach}, stochasticity can be introduced to a model through a distribution over the latent global parameters $\eta$. Examples of stochastic simulations are shown in Figure \ref{fig:exp:seir:stoch:full}. Clearly there is more capacity in this model for representing the underlying volatility and unpredictability of the precise nature of real-world phenomena, especially compared to the deterministic model.

However, this capacity comes with the reality that increased effort must be invested to ensure that the unknown latent states are correctly accounted for. For more specific details on dealing with this stochasticity please refer back to Section \ref{sec:approach}, but, in short, one must simulate for multiple stochastic values of the unknown parameters, for each value of the controllable parameters, and agglomerate the many individual simulations appropriately for the inference objective. When asking questions such as "will this parameter value violate the constraint?" there are feasibly some trajectories that are slightly above and some slightly below the trajectory generated by the deterministic simulation due to the inherent stochasticity (aleatoric uncertainty) in the real world. This uncertainty is integrated over in the stochastic model, and hence we can ask questions such as "what is the probability that this parameter will violate the constraint?" Using confidence values is this way provides some measure of how certain one can be about the conclusion drawn from the inference -- if the confidence value is very high then there is a measure of ``tolerance'' in the result, compared to a result with a much lower confidence. 

We define a joint distribution over model parameters as follows. We consider the $95\%$ confidence intervals of $\beta_1$, $\beta_2$, and $\beta_3$ and the values in the left hand-sides of Equations \eqref{eq:seir-quantities-begin}-\eqref{eq:seir-quantities-end-firstcolumn}, and assume that their true values are uniformly distributed across these confidence intervals. Then at each time $t$ in a simulation, we sample these values and then invert the system of Equations \ref{eq:seir-quantities-begin}-\ref{eq:seir-quantities-end} to obtain a sample of the model parameters. More sophisticated distributions could easily be introduced once this information becomes available. We now detail the nominal values used for typical trajectories (and the confidence intervals used for sampling). The nominal values are mostly the same as those used by~\citep{sei3r2020website}. We use: an incubation period of 5.1 days (4.5-5.8)~\citep{lauer2020incubation}; a mild infection duration of 6 days (5.5-6.5)~\citep{woelfel2020clinical}; a severe infection duration of 4.5 days (3.5-5.5)~\citep{sanche2020novel}; a critical infection duration of 6.7 days (4.2-10.4); fractions of mild, severe, and critical cases of $81\%$, $14\%$ and $5\%$~\citep{wu2020characteristics}; and a fatality ratio of $2\%$~\citep{wu2020characteristics}. We also use $\beta_1 = 0.33$ / day (0.23-0.43), and $\beta_2 = 0.$ / day (0.-0.05), and $\beta_3 = 0.$ / day (0.-0.025). Where possible, the confidence intervals are obtained from the studies which estimated the quantities. Where these are not given, we use a small range centred on the nominal value to account for possible imprecision.

\subsection{Agent-Based Simulation}
\label{sec:models:fred}

While compartmental models, such as the SEIR model described in Section \ref{sec:models:seir}, provide a mathematically well understood global approximation to disease dynamics, due to their coarse-grained statistical nature they cannot capture many important aspects and local details of the physical and social dynamics underlying the spread of a disease. These aspects include geographic information, spatio-temporal human interaction patterns in social hubs such as schools or workplaces, and the impact of individual beliefs on transmission events.
To address these limitations, agent-based simulators (ABS) have been introduced. Such simulators have practically no restrictions in terms of expressiveness, i.e., they can make use of all features of modern Turing-complete programming languages, at the significant computational cost of simulating all details involved.
\changed[Updates from Epistemix]{FRED\footnote{\url{https://fred.publichealth.pitt.edu/}} \cite{GRE-13-FRED} is an instance of the class of epidemiological agent-based simulators that are currently available for use in policy-making.  We chose to use FRED in this work because it has a version which is open-source and publicly-available.  FRED captures demographic and geographic heterogeneities of the population by modelling every individual in a region, including realistic households, workplaces and social networks. Using census-based models available for every state and county in the US and selected international locations, FRED simulates interactions within the population in discrete time steps of one day. Transmission kernels model the spatial interaction between infectious places and susceptible agents. Agents follow their own belief about susceptibility, severity and provided social benefits and barriers, such as health insurance and paid sick leave. Additionally, FRED can model immunity at the individual and population level, as well as the co-evolution of different viral strains. These characteristics enable FRED to provide much more fine-grained policy advice at either the regional or national level, based on socio-economic and political information which cannot be incorporated into compartmental models.

It is important to note here that FRED was built as a model for {\em influenza} pandemics, and in its publicly available version has not yet been adapted to \cov{}. Our experiments using FRED that appear later in Section \ref{sec:exp:fred} should be interpreted solely as illustrations of the methodology in Section \ref{sec:approach}, rather than as concrete results applying to the current \cov{} pandemic.
}{\subsubsection{FRED: Fine-grained simulation of disease spreading}
FRED\footnote{\url{https://fred.publichealth.pitt.edu/}} \cite{GRE-13-FRED} is an instance of the class of epidemiological agent-based simulators that are currently available for use in policy-making. FRED is an agent-based modeling language and execution platform for simulating changes in a population over time. FRED represents individual persons, along with social contacts and interactions with the environment. This enables the model to include individual responses and behaviors that vary according to the individual’s characteristics, including demographics (age, sex, race, etc.), as well as the individual’s interactions with members of various social interaction groups, such as their neighborhood, school or workplace. The FRED user can define and track any dynamic condition for the individuals within the population, including diseases (such as covid-19), attitudes (such as vaccine acceptance), and behaviors (such as social distancing).

FRED captures demographic and geographic heterogeneities of the population by modelling every individual in a region, including realistic households, workplaces and social networks. Using census-based models available for every state and county in the US and selected international locations, FRED simulates interactions within the population in discrete time steps of one hour. Transmission kernels model the spatial interaction between infectious places and susceptible agents. These characteristics enable FRED to provide much more fine-grained policy advice at either the regional or national level, based on socio-economic and political information which cannot be incorporated into compartmental models.

We chose to use FRED in this work because it has been used to evaluate potential responses to previous infectious disease epidemics, including vaccination policies \citep{lee2011benefits}, school closure \citep{potter2012school}, and the effects of population structure \citep{kumar2015population} and personal health behaviors \citep{kumar2013policies,liu2015role}.

After 10 years of develoment as an academic project, FRED has been licensed by the University of Pittsburgh to Epistemix \footnote{\url{http://www.epistemix.com/}}, to develop commercial applications of the FRED modeling technology.  In turn, Epistemix has developed a detailed COVID-19 model in FRED, which is used in the experiments described here.

The FRED COVID-19 model includes three interconnected components: (1) The Natural History of COVID-19; (2) The social dynamics/behavior of individuals; and (3) The Vaccination Program. The COVID-19 model was designed using the latest scientific data, survey information from local health authorities, and in consultation with expert epidemiologists. This model has been used to project COVID-19 cases in universities, K-12 school districts, large cities, and offices.

The FRED COVID-19 natural history model represents the period of time and trajectory of an individual from infection or onset to recovery or death. In the current version of the model, when an individual, or an agent, is exposed to SARS-CoV-2, the virus that causes COVID-19, the individual enters a 2-day latent period before they become infectious. In the infectious state, individuals can either be asymptomatic, symptomatic, or hospitalized. The probability of entering any of these infectious states is based on the individual’s age, infection history, and vaccination history. Individuals have a duration of illness (i.e. number of days they can transmit the virus) which is dependent on infectious state, a severity of disease (i.e. magnitude of transmissibility), and a disease outcome (recovery or death).

When an agent is exposed to SARS-CoV-2 and becomes symptomatic, the individual chooses whether or not to isolate themselves from normal activities. Approximately 20\% of individuals continue regular daily activities while symptomatic. Agents who are exposed and develop an asymptomatic infection do not isolate themselves and go about their regular activities. This introduces both symptomatic and asymptomatic forms of transmission into the model.

Prevalence of mask wearing and adherence to social distancing are unique to each location and change over time. The level of compliance to these behaviors is set based on the number of active infections that were generated from reported cases in the previous two weeks. Social distancing is assumed to reduce the number of contacts between agents in each place the agent attends.
}

Following the general recipe for framing planning as inference in Section \ref{sec:approach:abstract-model},
the following section defines what a prior on controls $\theta$ is in terms of of FRED internals, how FRED parameters relate to $\eta,$ and how to condition FRED on desirable future outcomes $Y^{aux}_{0:T}$.   
Section \ref{sec:models:fred:pyprob} describes results of performing automated inference in this stochastic simulation-based model using the probabilistic programming system PyProb.  The main point of this section is to illustrate how a stochastic simulator can be seen as a probabilistic model and, when integrated with an appropriate probabilistic programming system, can be repurposed to perform automatic inference, in this instance for planning as inference.  The FRED-specific recipe we provide below should be read with the understanding that it is easy to apply the same recipe to most if not all existing epidemiological simulators as their fundamental computational structure  is regular. 

\subsubsection{Turning FRED into A Probabilistic Program}
\label{sec:models:fred:probabilistic-model}

The FRED simulator has a parameter file which stipulates the values of $\theta$ and $\eta.$
In other words both the controllable and non-controllable parameters live in a parameter file.  FRED, when run given a particular parameter file, produces a sample from the distribution $p(X_{0:T},Z_{0:T}|\theta, \eta).$  Changing the random seed and re-running FRED will result in a new sample from this distribution.  

The difference between $X_{0:T}$ and $Z_{0:T}$ in FRED is largely in the eye of the beholder.  One way of thinking about it is that $X_{0:T}$ are all the values that are computed in a run and saved in an output file and $Z_{0:T}$ is everything else.  

In order to turn FRED into a probabilistic programming model useful for planning via inference several small but consequential changes must be made to it.  These changes can be directly examined by browsing one of the public source code repositories accompanying this paper.\footnote{\url{https://github.com/plai-group/FRED}}  First, the random number generator and all random variable samples must be identified so that they can be intercepted and controlled by PyProb.  Second, any variables that are determined to be controllable (i.e.~part of $\theta$) need to be identified and named.  Third, in the main stochastic simulation loop, the state variables required to compute $Y_t^{aux}$ and $Y_t^{obs}$ must be extracted.  Fourth, these variables must be given either synthetic ABC likelihoods or given constraints in the form of likelihoods.  Finally, a mechanism for identifying, recording, and or returning $X_{0:T}$ to the probabilistic programming system must be put in place.  FRED, like many stochastic simulators, includes the ability to write-out results of a run of the simulator to the filesystem.  This, provided that the correspondence between a sample $\theta^{(i)}$ and the output file or files that correspond to it is established and tracked, is how $X_{0:T}^{(i)}$ is implicitly defined.

In the interest of time and because we were familiar with the internals of PyProb and knew that we would not be using inference algorithms that were incompatible with this choice, the demonstration code does not show a full integration in which all random variables are controlled by the probabilistic programming system, instead, it only controls the sampling of $\theta$ and the observation of $Y_t^{aux}.$  Notably this means that inference algorithms like lightweight Metropolis Hastings \citep{wingate2011lightweight}, which are also included in PyProb, cannot be used with the released integration code.

\subsubsection{Details of FRED+PyProb Integration}\label{sec:models:fred:pyprob}

Our integration of PyProb into FRED required only minor modifications to FRED's code base, performed in collaboration with the FRED developed at Epistemx. More details about the integration of FRED and PyProb include:

\begin{enumerate}
    \item The simulator is connected to PyProb through a cross-platform execution protocol (PPX\footnote{\url{https://github.com/pyprob/ppx}}). This allows PyProb to control the stochasticity in the simulator, and requires FRED to wait, at the beginning of its execution, for a handshake with PyProb through a messaging layer.
    
    \item PyProb overwrites the policy parameter values $\theta$ with random draws from the user-defined prior. While PyProb internally keeps tracks of all random samples it generates, we also decided to write out the updated FRED parameters to a parameter file in order to make associating $\theta^{(i)}$ and $X_{0:T}^{(i)}$ easy and reproducible.
    
    \item For each daily iteration step in FRED's simulation, we call PyProb's \texttt{observe} function with a likelihood corresponding to the constraint we would like to hold in that day.
\end{enumerate}

\changed[R1.Q2.2]{With these connections established,  importance sampling of our inference objective in the FRED model can be directed by PyProb.}{With these connections established, we are able to select an inference engine implemented by PyProb to compute the posterior. We use a particularly simple algorithm, namely rejection sampling, in order to focus our exposition on the conceptual framework of planning as inference. PyProb implements multiple other, more complex, algorithms, which may be able to better approximate the posterior with a given computational budget. However for the inference task we consider, in which we attempt to infer only a small fraction of the random variables in the simulator, we find that rejection sampling is sufficiently performant.  }

We also remind the reader that, like in Section~\ref{sec:approach:time_varying_control}, more complex controls can be considered, in principle allowing for complex time-dependent policies to be inferred.  We do not examine this here, but note that this extension is straightforward to implement in the probabilistic programming framework, and that PyProb is particularly well adapted to coping with the additional complexity.
\changed[Update from Epistemix]{Compared to sampling parameter values for FRED at the beginning of the simulation, such time-varying policies are not readily available in FRED's configuration and to implement them would require changing FRED's internal state during the simulation.}{
Compared to sampling parameter values for FRED at the beginning of the simulation, such time-varying policies could be implemented through changing the FRED model source code directly. This approach will be explored in future research.
}

\section{Experiments}
\label{sec:exp:seir}

We now demonstrate how inference in epidemiological dynamics models can be used to inform policy-making decisions. We organize this section according to a reasonable succession of steps of increasing complexity that one might take when modelling a disease outbreak.  We again stress that we are not making \cov{} specific analyses here, but instead highlight how framing the task as in Section \ref{sec:approach} allows existing machine learning machinery to be leveraged to enhance analysis and evaluation of outcomes; and avoid some potential pitfalls.

We begin by showing how a simple, deterministic compartmental SEIR-based model can be used to inform policy-making decisions, and show how analysis derived from such a deterministic model can fail to achieve stated policy goals in practice. Next, we demonstrate how using a stochastic model can achieve more reliable outcomes by accounting for the uncertainty present in real world systems.  While these stochastic models address the limitations of the deterministic model, low-fidelity SEIR models are, in general, not of high enough fidelity to provide localized, region-specific policy recommendations. To address this we conclude by performing inference in an existing agent-based simulator of infectious disease spread and demonstrate automatic determination of necessary controls. 

\subsection{\fancyseir{} Model}
\label{sec:exp:fancyseir}
The most straightforward approach to modelling infectious diseases is to use low-dimensional, compartmental models such as the widely used susceptible-infectious-recovered (SIR) models, or the \fancyseir{} variant introduced in Section \ref{sec:models:seir}. These models are fast to simulate and easy to interpret, and hence form a powerful, low-overhead analysis tool. 

\subsubsection{Deterministic Model}
The system of equations defining the \fancyseir{} model form a deterministic system when global parameter values, such as the mortality rates or incubation periods, are provided. However, the precise values of these parameter values are unknown, and instead only confidence intervals for these parameters are known, i.e. the incubation period is between $4.5$ and $5.8$~\citep{lauer2020incubation}. This variation may be due to underlying aleatoric uncertainty prevalent in biological systems, or epistemic uncertainty due to the low-fidelity nature of SIR-like models. We do not discuss them here, but work exists automatically fitting point-wise estimates of model parameter values directly from observed data~\citep{wearing2005appropriate, mamo2015mathematical}. 

Regardless of whether one obtains a point estimate of the parameter values by averaging confidence intervals, or by performing parameter optimization, the first step is to use these values to perform fully deterministic simulations, yielding simulations such as those shown in Figure \ref{fig:exp:seir:det_nom}. Simulations such as this are invaluable for understanding the bulk dynamics of systems, investigating the influence of variations in global parameter values or investigating how controls affect the system. However, the ultimate utility in these models is to \emph{use} them to inform policy decisions to reduce the impact of outbreaks. As eluded to above, this is the primary thrust of this work, combining epidemiological simulators with automated machine learning methodologies to model policy outcomes, by considering this problem as \emph{conditioning} simulations on outcomes. 

To demonstrate such an objective, we consider maintaining the infected population below a critical threshold $C$ at all times. In a deterministic system there are no stochastic quantities and hence whether the threshold is exceeded is a deterministic function of the controlled parameters, i.e. the value of $p(\forall_{t > 0} Y_t^{aux}=1 | \theta)$ (related to \eqref{eq:control-inference} via Bayes rule) is binary in a deterministic system and hence takes a value of either $0$ or $1$.  Therefore, we can simply simulate the deterministic system for a finite number of $\theta$ values, and select those parameter values that do not violate the constraint.  We vary the free parameter $u\in\left[0, 1\right]$, where $u$ is a scalar value that reduces the baseline reproduction rate as $\hat{R}_0 = (1-u)R_0$.  We define $u$ in this way such that $u$ represents an \emph{intervention}, or change from normal conditions. The parameter $u$ is the only parameter we have control over, and hence $\theta = u$.

\begin{figure}[t]
    \centering

    \begin{subfigure}[b]{0.48\textwidth}
        \centering
        \includegraphics[width=\textwidth]{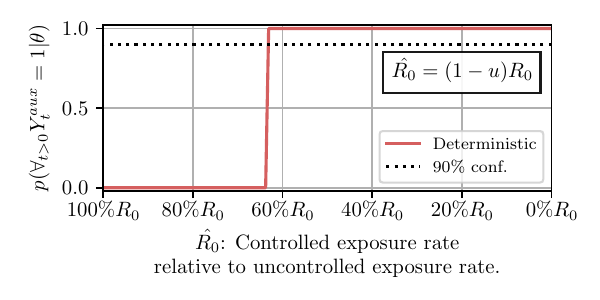}
        \caption{ }
        \label{fig:exp:seir:det_plan:det_param}
    \end{subfigure}
    ~
    \begin{subfigure}[b]{0.48\textwidth}
        \centering
        \includegraphics[width=\textwidth]{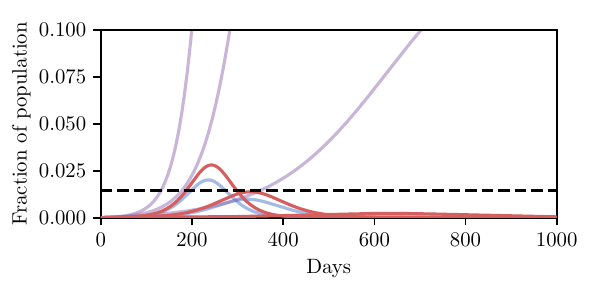}
        \caption{}
        \label{fig:exp:seir:det_plan:det_traj}
    \end{subfigure}
    
    \caption{Here we demonstrate planning using the deterministic \fancyseir{} model. 
    Figure \ref{fig:exp:seir:det_plan:det_param} shows, in red, the probability that the 
    constraint is met using the deterministic simulator. The probability jumps from zero
    to one at a value of approximately $u=0.37$.
    Figure \ref{fig:exp:seir:det_plan:det_traj} then shows trajectories using three 
    salient parameter values, specifically, $u=\left[ 0.3, 0.37, 0.45 \right]$.
    Each parameter value corresponds to more control than the previous, and is evident
    as the peak infection fraction drops from approximately $0.03$ with $u=0.3$, to 
    $0.014$ with $u=0.37$ and almost zero with $u=0.45$. These values were selected
    as values just below, on, and above the threshold seen in Figure 
    \ref{fig:exp:seir:det_plan:det_param}. We reuse the colour scheme defined in 
    Figure \ref{fig:exp:seir:det_nom}.
    }
    \label{fig:exp:seir:det_plan}
\end{figure}

Results for this are shown in Figure \ref{fig:exp:seir:det_plan}.  It can then be read off that under the deterministic model $\hat{R}_0$ must be reduced by at least $37.5\%$ of $R_0$ to satisfy the constraint. Figure \ref{fig:exp:seir:det_plan} shows trajectories simulated using insufficient intervention with $u=0.3$ ($\hat{R}_0 = 70\%R_0$), acceptable intervention of $u=0.375$ ($\hat{R}_0 = 62.5\%R_0$), and excessive intervention of $u=0.45$ ($\hat{R}_0 = 55\%R_0$), and show that these parameters behave as expected, violating the constraint, remaining just under the threshold and remaining well beneath the threshold 
respectively.

\subsubsection{Stochastic Simulation}
While the above example demonstrates how parameters can be selected by conditioning on desired outcomes, we implicitly made a critical modelling assumption. While varying the free parameter $u$, we fixed the \emph{other} model parameter values ($\alpha^{-1}, \gamma_1$, etc) to single values. We therefore found a policy intervention in an unrealistic scenario, namely one in which we (implicitly) claim to have certainty in all model parameters except $u$.

To demonstrate the pitfalls of analyzing deterministic systems and applying the results to an inherently stochastic system such as an epidemic, we use the permissible value of $u$ solved for in the deterministic system, $u=0.375$, and randomly sample values of the remaining simulation parameters. This ``stochastic'' simulator is a more realistic scenario than the deterministic variant, as each randomly sampled $\eta$ represents a unique, plausible epidemic being rolled out from the current world state.

\begin{figure}[tb]
    \centering

    \begin{subfigure}[t]{\textwidth}
        
        \begin{subfigure}[t]{0.46\textwidth}
            \centering
            \includegraphics[width=\textwidth]{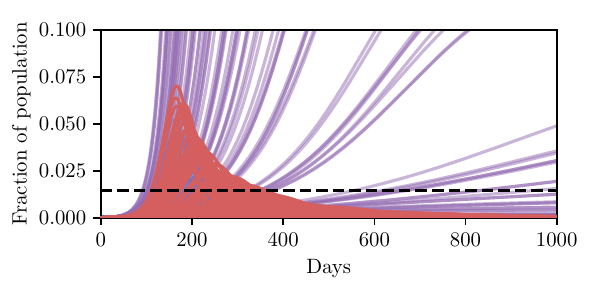}
            \caption{$\hat{R}_0=0.63R_0$}
            \label{fig:exp:seir:stoch_det:under}
        \end{subfigure}
        ~
        \begin{subfigure}[t]{0.46\textwidth}
            \centering
            \includegraphics[width=\textwidth]{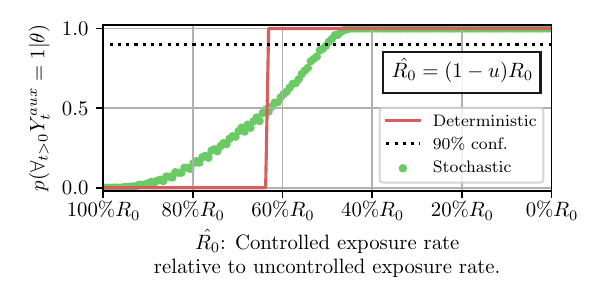}
            \caption{ }
            \label{fig:exp:seir:stoch_det:stoch_param}
        \end{subfigure}

    \end{subfigure}

    \begin{subfigure}[b]{\textwidth}
        
         \begin{subfigure}[b]{0.46\textwidth}

            \begin{subfigure}[b]{\textwidth}
                \centering
                \includegraphics[width=\textwidth]{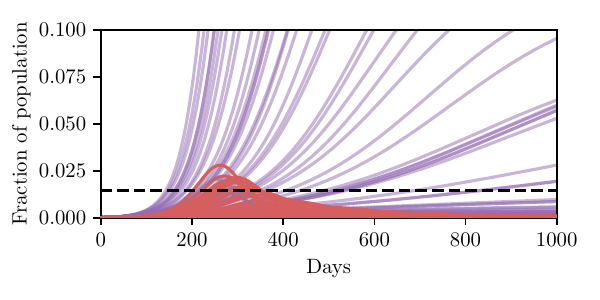}
                \caption{$\hat{R}_0=0.5R_0$}
                \label{fig:exp:seir:stoch_det:border}
            \end{subfigure}
            
             \begin{subfigure}[b]{\textwidth}
                \centering
                \includegraphics[width=\textwidth]{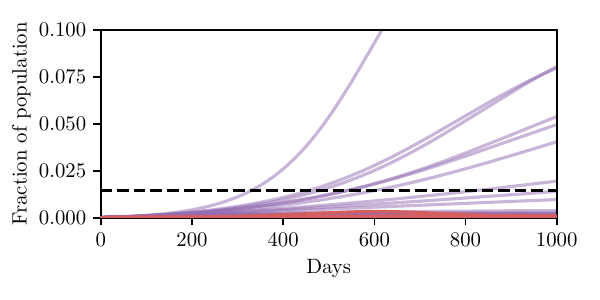}
                \caption{$\hat{R}_0=0.4R_0$}
                \label{fig:exp:seir:stoch_det:safe}
            \end{subfigure}
            
        \end{subfigure}
        ~
        \begin{subfigure}[b]{0.48\textwidth}
            \includegraphics[width=\textwidth]{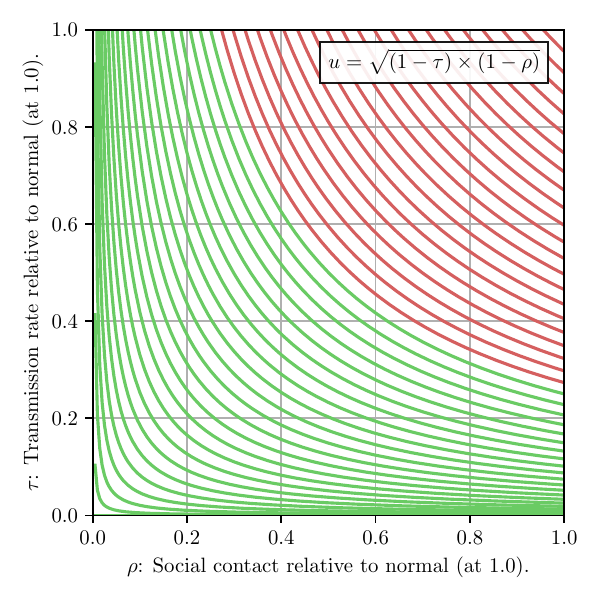}
            \centering
            \caption{Green = valid, red = invalid.}
            \label{fig:exp:seir:stoch_det:policy}
        \end{subfigure}
        
    \end{subfigure}

    \caption{Comparison of stochastic and deterministic \fancyseir{} models for policy selection. We reuse the colour scheme defined in 
             Figure \ref{fig:exp:seir:det_nom} for trajectories. %
             Figure \ref{fig:exp:seir:stoch_det:under} shows a stochastic simulation using $\hat{R}_0=0.63R_0$,
             identified as an acceptable parameter value under the deterministic model. However, once used
             in a stochastic system, the parameter performs poorly, yielding many simulations that violate
             the constraint. This highlights why analyses in deterministic systems can yield poor results. %
             Figure \ref{fig:exp:seir:stoch_det:stoch_param} repeats the analysis in 
             Figure \ref{fig:exp:seir:det_plan:det_param} adding planning in the stochastic simulation,
             where the $y$-coordinate can now be interpreted as the confidence level. %
             Figure \ref{fig:exp:seir:stoch_det:border} shows a simulation using the lowest valid value of $u$,
             representing the ``weakest'' valid policy. This value, approximately $\hat{R}_0=0.5R_0$,  
             renders most of the trajectories under the threshold, with only a small fraction above, implying 
             that it will satisfy the criteria with high probability. %
             Figure \ref{fig:exp:seir:stoch_det:safe} shows simulations using $\hat{R}_0=0.4R_0$ effectively 
             reduces the level of infection to near zero. 
             Figure \ref{fig:exp:seir:stoch_det:policy} illustrates an example of how policy-level variables 
             create model-level parameter values. Shown are the level-sets of the free parameter $u\in\left[ 0, 1 \right]$, which acts to reduce the baseline reproduction rate $R_0$ as $\hat{R}_0 = (1-u) R_0$. 
             We suggest that the reduction in the (unknown) reproduction rate is given by the root
             of the product of two factors controllable through policy. Green level sets 
             indicate that the value of $u$ was effective and achieved a $90\%$ confidence that the trajectory
             does not violate the constraint, whereas red curves do not satisfy this. 
             }
    \label{fig:exp:seir:planning}
\end{figure}

The results are shown in Figure~\ref{fig:exp:seir:stoch_det:under}. Each line represents a possible epidemic. We can see that using the previously found value of $u$ results in a large number of epidemics where the infectious population exceeds the constraint, represented by the red trajectories overshooting the dotted line. Simply put, the control parameter we found previously fails in as unacceptable number of simulations.

This detail highlights the shortcomings of the deterministic model: in the deterministic model a parameter value was either accepted or rejected with certainty. There was no notion of the variability in outcomes, and hence we have no mechanism to concretely evaluate the risk of a particular configuration.

Instead, we can use a stochastic model which at least does account for some aleatoric uncertainty about the world. We repeat the analysis picking the required value of $u$, but this time using the stochastic model detailed in Section~\ref{sec:models:seir}. In practice, this means the (previously deterministic) model parameters detailed in Equations~\ref{eq:seir-quantities-begin} - \ref{eq:seir-quantities-end} are randomly sampled for each simulation according to the procedure outlined following the equations. 

To estimate the value of $p\left(\forall_t : Y_t^{aux} = 1 | \theta \right)$, for a given $u$ value, we sample $M$ stochastic trajectories from the system. We then simply count the number of trajectories for which the condition $\forall_t : Y_t^{aux} = 1$ holds, and divide this count by $M$.  Intuitively, this is operation is simple: for a given $\theta$, simulate a number of possible trajectories, and, as the number of simulations $M$ tends to infinity, the fraction that satisfy the constraint is the desired probability value. We note that this operation corresponds to an ``inner'' Monte Carlo expectation, sampling under the distribution of simulator trajectories conditioned on $\theta$, evaluating the expected number of trajectories that do not violate the threshold. This value is then passed through a non-linear indicator function extracting those parameters that yield a confidence above a certain threshold. We are then free to use any method we please for exploring $\theta$ space, or, evaluating additional Monte Carlo expectations under the resulting $\theta$ distribution. As such, this system is a nested Monte Carlo sampler~\cite{rainforth2017nesting}. 

The results are shown in Figure~\ref{fig:exp:seir:stoch_det:stoch_param}.  The certainty in the result under the stochastic model is not a binary value like in the deterministic case, and instead occupies a continuum of values representing the confidence of the results. We see that the intersection between the red and green curves occurs at approximately $0.5$, explaining the observation that approximately half of the simulations in Figure \ref{fig:exp:seir:stoch_det:under} exceed the threshold. We can now ask questions such as: "what is the parameter value that results in the the constraint not being violated, with $90\%$ confidence." We can read off rapidly that we must instead reduce the value of $\hat{R}_0$ to $50\%$ of its original value to satisfy this confidence based constraint.  Repeating the stochastic simulations using these computed values confirms that very few simulations violate the constraint (Figure \ref{fig:exp:seir:stoch_det:border}). The ability to tune the outcome based on a required level of confidence is paramount for safety-critical applications, as it informs how sensitive the system is to the particular parameter choice and is more resilient to model misspecification.

\subsubsection{Model predictive control}
We have shown how one can select the required parameter values to achieve a desired objective. To conclude this example, we apply the methodology to iterative planning. The principal idea underlying this is that policies are not static and can be varied over time conditioned on the current observed state.  \changed[R1.5.3, R1.5.minor.3]{Under the formalism used here, this is as simple as re-applying the stochastic planning each time step to produce a new policy conditioned on new information. }{Under the formalism used here, re-evaluating the optimal control to be applied, conditioned on the new information, is as simple as re-applying the planning algorithm at each time step.  Note that here we consider constant control.  However, more complex, time-varying control policies can easily be considered under this framework.  For instance, instead of recovering a fixed control parameter, the parameter values of a polynomial function defining time-varying control input could be recovered, or, a scalar value determining the instantaneous control input at each time-step.  This is a benefit of the fully Bayesian and probabilistic programming-based approach we have taken: the model class that can be analyzed is not fixed and can be determined (and easily changed and iterated on) by the modeller, and the inference back-end cleanly and efficiently handles the inference.   }

\begin{figure}[t]
    \centering

    \begin{subfigure}[b]{0.48\textwidth}
        \centering
        \includegraphics[width=\textwidth]{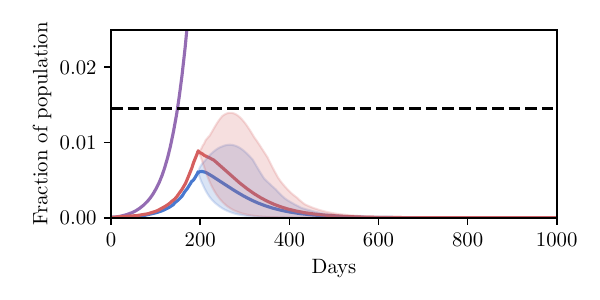}
        \caption{ }
        \label{fig:exp:seir:mpc:t1:zoom}
    \end{subfigure}
    ~
    \begin{subfigure}[b]{0.48\textwidth}
        \centering
        \includegraphics[width=\textwidth]{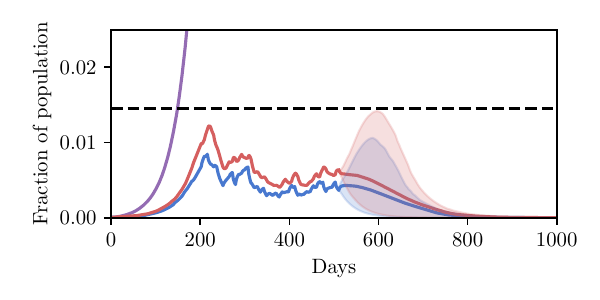}
        \caption{ }
        \label{fig:exp:seir:mpc:t2:zoom}
    \end{subfigure}
    
    \caption{Here we briefly demonstrate the capacity of the \fancyseir{} model in a model predictive control setting. Figure \ref{fig:exp:seir:mpc:t1:zoom} shows the state when we begin controlling the system  at $t=200$ with some level of infection already present. We solve for the minimum required control such that the constraint is satisfied. We plot the $90\%$ confidence interval over trajectories conditioned on this control value. We then step through the system, randomly sampling continuations, and adapting the controls used such that the constraint is always met (Figure \ref{fig:exp:seir:mpc:t2:zoom}). We uncover that stronger controls must be applied early on to reduce the infected population, but that the amount of control required can then reduce over time as herd immunity becomes a stronger effect.  We reuse the color scheme defined in Figure \ref{fig:exp:seir:det_nom} for plotting trajectories. }
    \label{fig:exp:seir:mpc}
\end{figure}

We show a demonstration of this in Figure \ref{fig:exp:seir:mpc}. In this example, we begin at time $t=200$ with non-zero infection rates. We solve for a policy that satisfies the policy with $90\%$ certainty, and show this confidence interval over trajectories as a shaded region. We then simulate the true evolution of the system for a single step sampling from the conditional distribution over state under the selected control parameter. We then repeat this process at regular intervals, iteratively adapting the control to the new world state.  We see that the confidence criterion is always satisfied and that the infection is able to be maintained at a reasonable level. We do not discuss this example in more detail, and only include it as an example of the utility of framing the problem as we have, insomuch as iterative re-planning based on new information is a trivial extension under the formulation used.

\subsubsection{Policy-based controls}
We have illustrated how simulations can be used to answer questions about the suitability of parameter values we can influence, while marginalizing over those parameter we do not have control over. However, $u$ is not something that is \emph{directly} within our control.  Instead, the value of $u$ is set through changing policy level factors. As an exploratory example we suggest that the value of $u$ is the square root of the product of two policy-influenceable factors: the fractional reduction in social contact, $\rho$, below its normal level (indicated as a value of $1.0$), and the transmission rate relative to the normal level, $\tau$, where we again denote normal levels as $1.0$. This relationship is shown in Figure \ref{fig:exp:seir:stoch_det:policy}.

We indicate $u$ level sets that violate the constraint in red, and valid sets in green. We suggest taking the least invasive, valid policy, being represented by the highest green curve. Once the analysis above has been performed to obtain a value of $u$, that satisfies the required infection threshold, it defines the set of achievable policies.  Any combination of $\tau$ and $\eta$ along this curve render the policy valid. Here, additional factors may come into consideration that make particular settings of $\tau$ and $\eta$ more or less advantageous. For instance, wearing more PPE may be cheaper to implement and less economically and socially disruptive than social distancing, and so higher values of $\tau$ may be selected relative to $\eta$. This reduces to a simple one-dimensional optimization of the cost surface along the level-set.

While we have simply hypothesized this as a potential relationship, it demonstrates how policy level factors influence simulations and outcomes. While the SEIR model family is an invaluable tool for analyzing and understanding the bulk dynamics of outbreaks, it is too coarse-grained for actual, meaningful, localized policy decisions to be made, especially when those policy decisions are directly influencing populations. Further, these notions of ``policy'' are somewhat abstract here because of the high-level nature of the \fancyseir{} model used. We now go on to resolve these issues by using the more sophisticated, agent-based simulator, FRED, where simulations are able to represent localized variations, and, where real policy measures are more easily defined.

\subsection{FRED Simulator}
\label{sec:exp:fred}

In this section, we turn to agent-based simulators. We showcase how control as inference might possibly be used to inform \textit{regional} policy decisions through two scenarios presented in Sections~\ref{sec:exp:fred:influenza},\ref{sec:exp:fred:covid}. In the first scenario, we consider an \emph{influenza} outbreak and in the second a \cov{} outbreak is simulated. In both scenarios the policy makers wish to ``flatten the curve'' by limiting some statistic of the infected population under a certain threshold.

\begin{figure}
    \centering
    
    \begin{subfigure}[b]{0.48\textwidth}
        \centering
        \includegraphics[width=\textwidth]{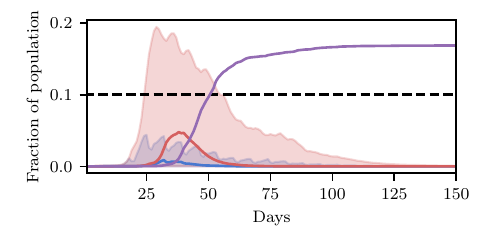}
        
        \includegraphics[width=\textwidth]{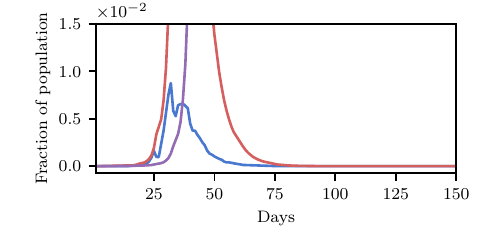}
        \caption{Without controls.}
        \label{fig:FRED-SEIR-summaries:uncontrolled}
    \end{subfigure}%
    \hfill%
    \begin{subfigure}[b]{0.48\textwidth}
        \centering
        \includegraphics[width=\textwidth]{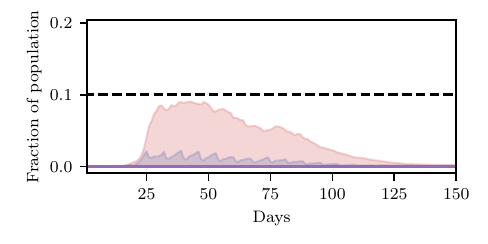}
        
        \includegraphics[width=\textwidth]{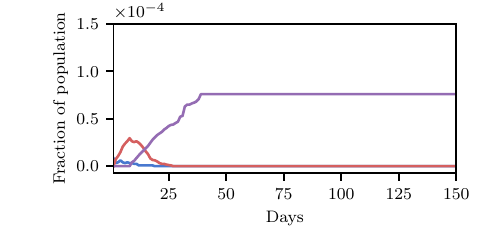}
        \caption{With controls.}
        \label{fig:FRED-SEIR-summaries:uncontrolled}
    \end{subfigure}
    \caption{\changed[R3 Q3]{SEIR statistics extracted from a FRED simulation of an \emph{influenza} (not \cov{}) outbreak in Allegheny County.  (left) controlled scenarios keeping the number of infectious people below $10\%$ (black dotted line), and (right) uncontrolled scenarios. We plot the median and confidence bands between 3rd and 97th percentile. On the left the confidence interval for infectious people (red) stays below our constraint, verifying that all controlled runs follow our policy, while the confidence band for the uncontrolled scenarios violates our policy constraint. The effect is also visible for the median run which spikes much stronger under uncontrolled policy. The susceptible statistic (S) is left out here to focus on the relevant dynamics.}{Aggregated SEIR statistics extracted from $100,000$ FRED simulations of an \emph{influenza} (not \cov{}) outbreak in Allegheny County.  The objective in this scenario is to keep the number of infectious people (I in SEIR terminology, shown here in red) below $10\%$, indicated by the black dotted line.  We plot the median and confidence intervals between $3\textsuperscript{rd}$ and $97\textsuperscript{th}$ percentile, shown by the shaded areas.  To avoid clutter and focus on relevant dynamics, the susceptible statistic (S in SEIR terminology, shown previously in green) and confidence intervals for the Recovered (R in SEIR terminology, shown here in purple) statistic are omitted.  Blue shows the fraction of the population that has been exposed (E in SEIR terminology).  The bottom row shows a zoomed-in version of the top row, where all confidence intervals have been removed, for ease of visual inspection.  Figure \ref{fig:FRED-SEIR-summaries:uncontrolled} shows the evolution of the outbreak when no controls are applied.  Figure \ref{fig:FRED-SEIR-summaries:uncontrolled} shows the evolution of the outbreak when controls, solved for using our method, are applied.  When no controls are applied we see the number of infectious people often exceeds the allowable threshold, whereas, when optimal controls are applied, the confidence interval for infectious people (red) stays below our constraint. }}
    \label{fig:FRED-SEIR-summaries}
\end{figure}

\subsubsection{Influenza simulation}
\label{sec:exp:fred:influenza}
In this section, we consider a scenario where an \emph{influenza} (not \cov{}) outbreak has occurred in Allegheny County (similarly to \cite{GRE-13-FRED}), and policy makers wish to limit the number of infected to less than 10\% of the county’s population.
To achieve this hypothetical goal policy-makers might consider the following five controls among others (corresponding to the $\theta$ parameter defined in Table~\ref{tab:prior_params}):
\begin{itemize}
    \item A ``social distancing'' policy which mandates all citizens stay home for a fixed period of time. Here, policy makers must influence:
    \begin{enumerate}
        \item $\theta_1,$ a \emph{shelter-in-place duration}, or the length of time a social distancing policy must be in place.
        \item $\theta_2,$ a \emph{shelter-in-place compliance rate}, or the percentage of the population to which this policy applies.
    \end{enumerate}
    \item $\theta_3,$ a \emph{symptomatic isolation rate}, the fraction of symptomatic individuals that self isolate during an epidemic. 
    \item $\theta_4,$ a \emph{school closure attack rate threshold}, a threshold on the total percentage of people infected that automatically triggers a three-week school closure when exceeded.  
    \item $\theta_5,$ a \emph{hand washing compliance}, the percentage of the population that washes their hands regularly.
\end{itemize}

\newcommand{\unif}[2]{\text{Uniform}(#1, #2)}

For simplicity of results interpretation we have put uniform priors, appropriately continuous or discrete, on intervals of interest for these controllable parameters.

Also relative to \eqref{eq:control-inference} we choose $Y_t$ to be a binary variable indicating if the proportion of the county's infected population is below 10\% on day $t$. By inferring $p(\theta | \forall_t : Y_t = 1$), we characterize which control values will lead to this desired outcome.

\begin{table}[t]  
    \centering
\begin{tabular}{l|l}
     \textbf{Prior} & \textbf{Control} \\
     \hline
      $\theta_1 \sim \, $ \unif{0}{14} & \textit{shelter in place duration}    \\
      $\theta_2  \sim\, $ \unif{0}{1} & \textit{shelter in place compliance rate}   \\
      $\theta_3  \sim\, $  \unif{0}{1} & \textit{isolation rate}  \\
      $\theta_4  \sim\, $ \unif{0.01}{0.21} & \textit{school closure attack rate threshold}  \\
      $\theta_5  \sim\, $ \unif{0}{1} & \textit{hand washing compliance rate}   \\
     \hline
\end{tabular}
    \caption{Prior over FRED control parameter $\theta = \{\theta_1, \ldots, \theta_5\}$ for Influenza simulations.}
    \label{tab:prior_params}
\end{table}

Using the model described in Section~\ref{sec:models:fred} with parameters as defined in this section, we used PyProb to perform automated inference over the policy parameters described in Table \ref{tab:prior_params}.  To reiterate we infer the posterior distribution over control parameters $\theta$ which satisfy the goal of limiting the instantaneous number of infected to less than 10\% of the county's population at all times up to a maximum of 150 days.  
In conducting our experiments we generated one million samples, 40\% of which satisfied the policy goal. 

\paragraph{Simulator configuration}
We simulate Allegheny County, Pennsylvania, using 2010 U.S. Synthetic Population data.
\changed[]{We use the default FRED parameters,}{
We use an older, publicly available version of FRED\footnote{\url{https://github.com/PublicHealthDynamicsLab/FRED}} and its default parameters,
}
which simulates a model of influenza ``in which infectivity starts 1 day before symptoms and lasts for 7 days.  Both symptoms and infectivity ramp up and ramp down.''\footnote{\url{https://github.com/PublicHealthDynamicsLab/FRED/blob/FRED-v2.12.0/input_files/defaults}} All simulations begin with 10 random agents seeded with the virus at day zero. Person-to-person contact rates for various environments (household, office, etc) were calibrated to specifically model Allegheny County. Parameter files for the four policies under consideration are available online.\footnote{\url{https://github.com/plai-group/FRED/tree/FRED-v2.12.0/params}} \changed[]{Unfortunately COVID-19 parameter files for FRED were not available at the time of writing.}{}

\begin{figure}
    \centering
    \includegraphics[width=0.9\textwidth]{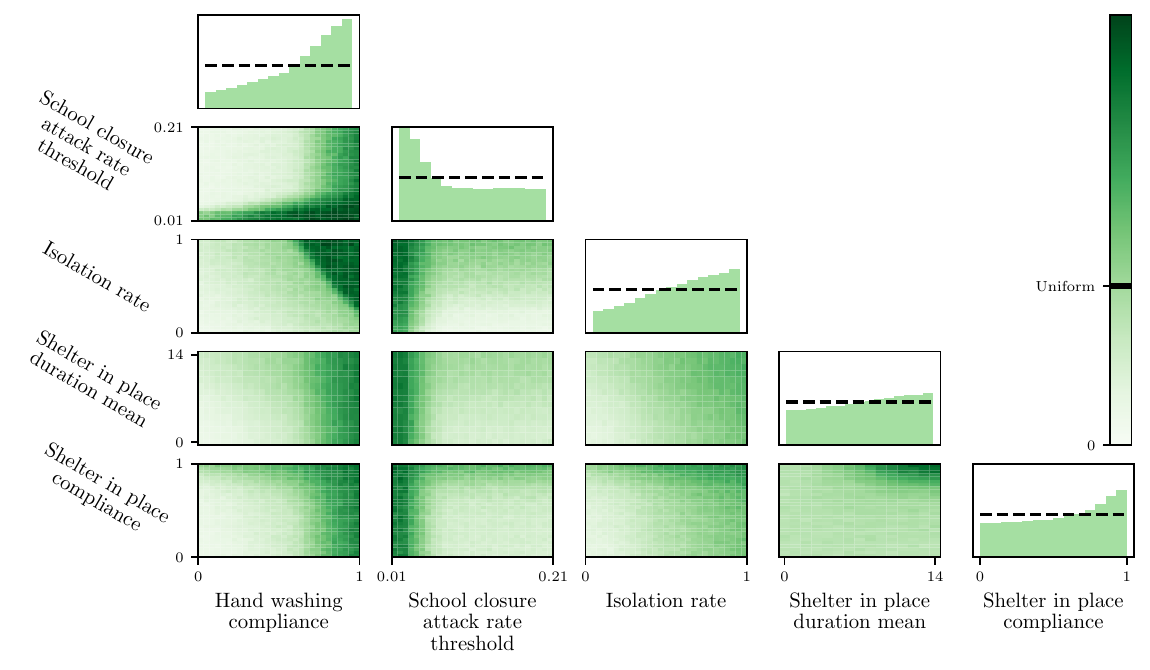}
    \caption{\changed[R3 Q3]{Array of 2D histograms showing two-dimensional marginal distributions over controllable policy parameters that give rise to appropriately controlled outcomes in Allegheny county.  Marginals for each policy are shown in the bottom row, with the number of samples from the uniform prior indicated by the dashed line. We can clearly see the efficacy of high rates of hand washing and a quick school closure policy, as indicated by the non-uniformity of the marginal distributions.}{Empirically determined marginals of the full joint distribution over controllable policy parameters that lead to desired outcomes (see Equation \eqref{eq:control-inference}) for a simulated influenza epidemic in Allegheny county.  Along the diagonal: one-dimensional marginals of each control parameter with a uniform prior indicated by the dashed line.  We can clearly see the efficacy of high rates of hand washing and a quick school closure policy, as indicated by the non-uniformity of the marginal distributions. The remaining array of plots show two-dimensional marginal distributions of any two parameters where the dark green color indicates higher probability density.  For reference, the color corresponding to a uniform prior on the two-dimensional plots is indicated on the color bar on the right.  This illustrates policy-level outcomes such as there being a strong interaction when jointly enforcing high \emph{isolation rate} and \emph{hand washing compliance}.  In contrast, the effectiveness of \emph{school closure attack rate threshold} is largely independent of the \emph{isolation rate} (and indeed most other parameters). Such richly structured information is paramount for making effective and  \emph{justifiable} policy decisions, and is only provided through fully Bayesian analysis, as opposed to simpler reinforcement learning or optimal control methodologies which may only provide a point estimate of the optimal control to be applied, with no quantification of uncertainty or the joint interaction of parameters.  } }
    \label{fig:FRED-scatter-plot-grid}
\end{figure}

\paragraph{Results}
Figure~\ref{fig:FRED-SEIR-summaries} shows the SEIR statistics for both the controlled (left) and uncontrolled (right) simulations. We see that the confidence interval in red stays beneath our infection rate constraint indicated by the dashed black line, while in the uncontrolled scenario the confidence band well exceeds the constraint. We also observed the overall number of exposed and infectious populations is much lower using samples of control variables from the inferred posterior, which indicates that control variable values draw from this posterior are indeed effective in limiting the spread of the virus. 

\begin{figure}
    \centering
    \begin{subfigure}[b]{\textwidth}
        \centering
        \includegraphics[width=0.5\linewidth]{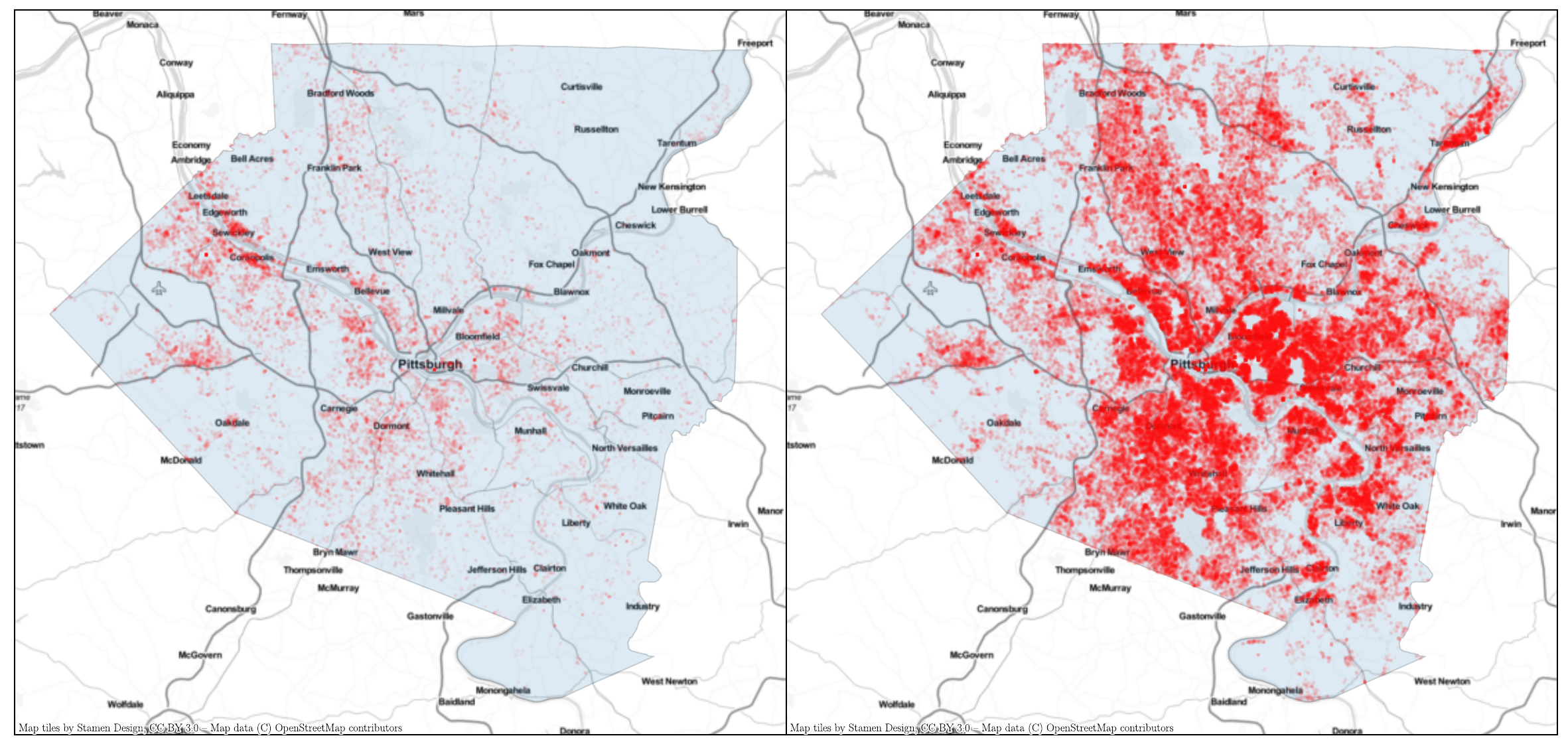}
        \caption{Day 30: controlled = 15,626, uncontrolled = 208,646}
    \end{subfigure}
    \begin{subfigure}[b]{\textwidth}
        \centering
        \includegraphics[width=0.5\linewidth]{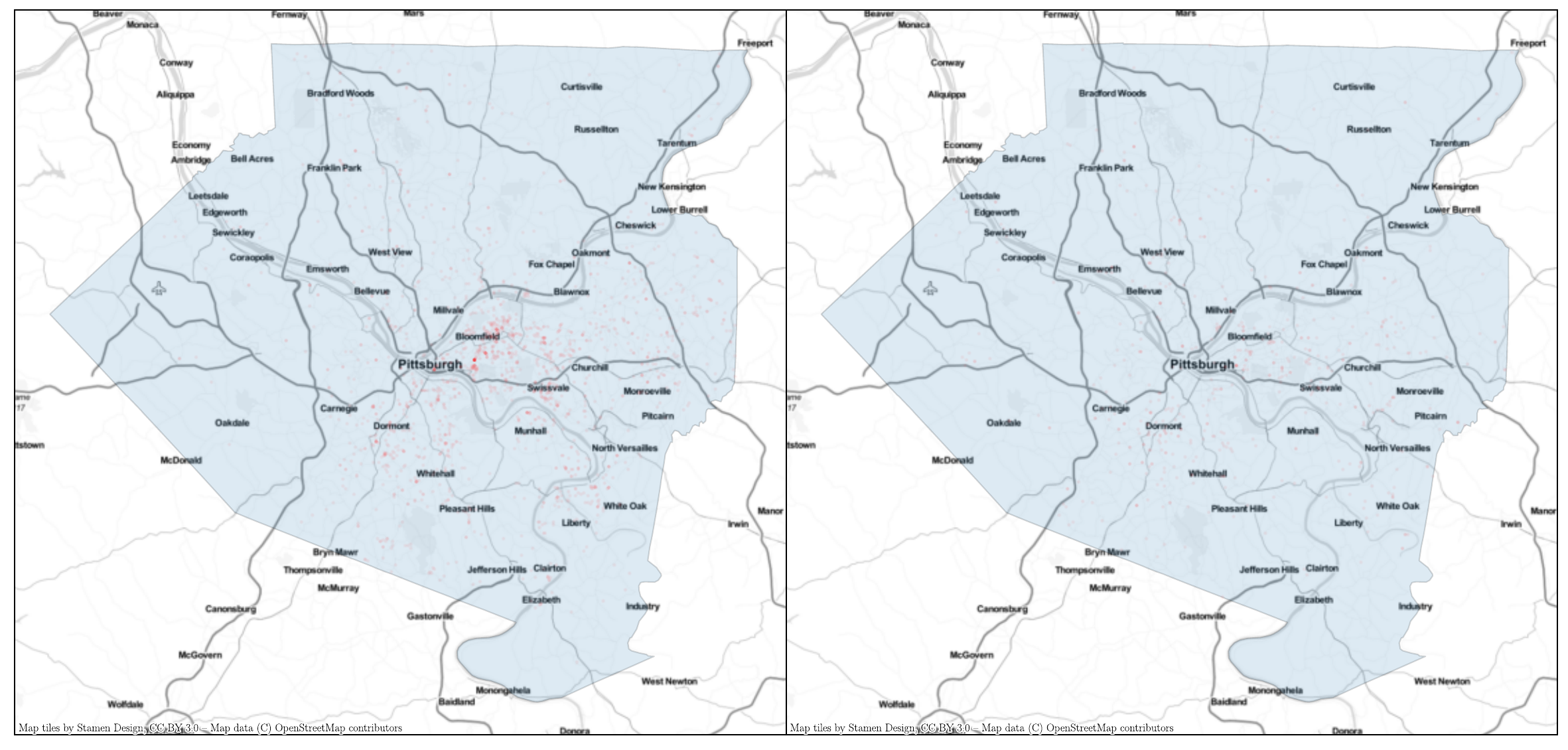}
                \caption{Day 60: controlled = 1,567 uncontrolled = 530}
        \begin{subfigure}[b]{\textwidth}
            \centering
            \includegraphics[width=0.5\linewidth]{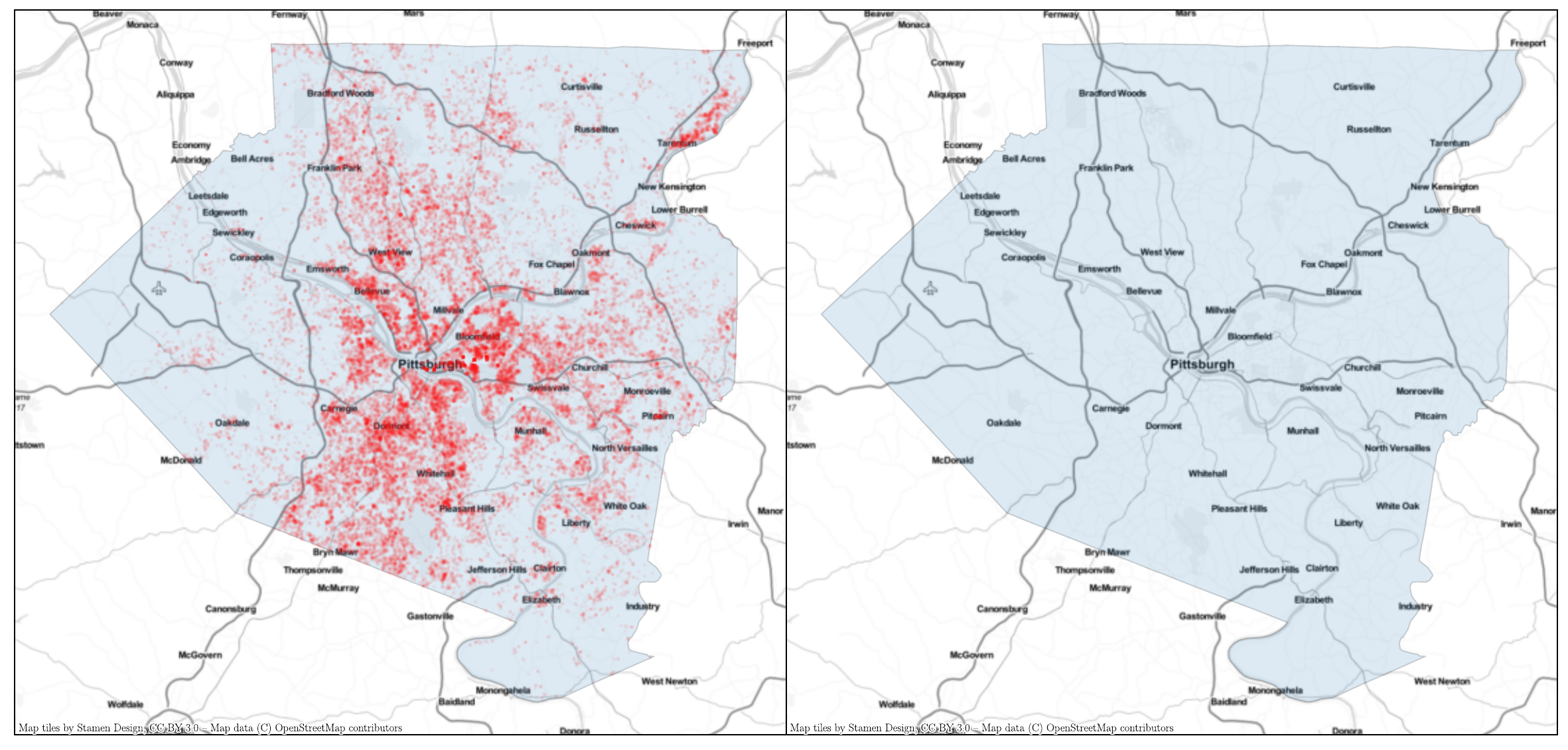}
            \caption{Day 90: controlled = 42,133, uncontrolled = 0}
            \label{fig:FRED-allegheny-county-over-time-90}
        \end{subfigure}
    \begin{subfigure}[b]{\textwidth}
        \centering
        \includegraphics[width=0.5\linewidth]{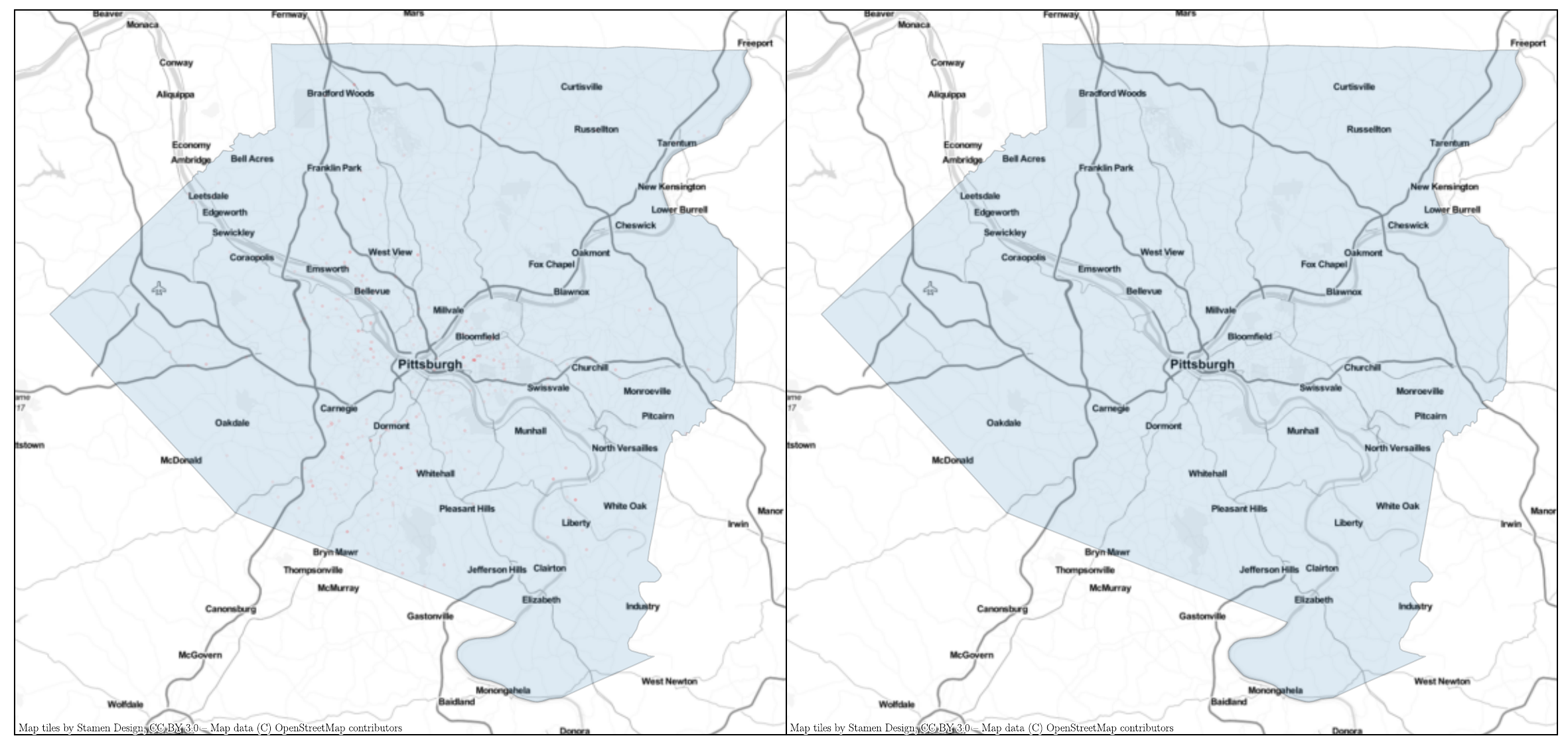}
        \caption{Day 120: controlled = 1,783, uncontrolled = 0}
    \end{subfigure}
    \end{subfigure}
    \caption{Progression of a simulated influenza epidemic in Allegheny county under controlled (left) and uncontrolled (right) scenarios. Each red dot represents the household of an infectious person. The count of the number of infected households in each scenario appears in the captions below each row. The peak number of cases in the uncontrolled scenario is $215,799$ on day $29$, while the peak number of cases in the controlled scenario is $65,997$ on day $83$.  \changed{}{We see that the controls we solve for successfully ``flatten the curve,'' indicated by a much lower density of red dots.  There is also a second spike predicted, visible in Figure \ref{fig:FRED-allegheny-county-over-time-90}, where as controls are removed there is an increase in cases throughout the susceptible portion of the population.  However, this second spike is still below the required threshold. }}
    \label{fig:FRED-allegheny-county-over-time}
\end{figure}

While Figure~\ref{fig:FRED-SEIR-summaries} indicates that the inferred controls can achieve the desired aims, it doesn't indicate \textit{which} policy to choose. To answer this, we plot the two-dimensional marginal distributions over controlled policy parameters in Figure~\ref{fig:FRED-scatter-plot-grid}. Policy-makers could use a figure like this, coupled with a utility function, to generate policy recommendations.  Interpreting Figure~\ref{fig:FRED-scatter-plot-grid}, first we see the importance of influencing hand washing compliance and quick school closures.  This plot indicates that hand washing compliance must be driven above 50\% and the school closure attack rate threshold beneath ~7\% in order to achieve our stated goal. We also note the correlation between hand washing compliance and isolation rate.  If we can achieve only 70\% hand washing compliance the isolation rate must be driven high, however, if hand washing compliance is very good then a lower isolation rate is tolerable. The importance of hygiene and long-term school closures has also been noted in the epidemiology literature \cite{germann2019school, saunders2017effectiveness, lee2010simulating}.

A final interpretation of the results of Figure~\ref{fig:FRED-scatter-plot-grid} can be provided in terms of the outcome of a hypothetical \emph{influenza} policy-making scenario.  A recommendation one might extract from this visualization of the posterior inference results is a conjunction of the following controls:
\begin{enumerate}
    \item Ensure that all schools close as soon as 2\% of the population contracts the virus.
    \item Attempt to drive the hand washing compliance to 80\%.
    \item Attempt to drive the symptomatic isolation rate to 50\%.
    \item No amount of sheltering in place is required.
\end{enumerate}

Such a recommendation corresponds to what one could imagine would be, in comparison to more draconian options, a relatively mild, economically speaking, policy response that still attains the objective. Of course, in the presence of a real influenza outbreak, vaccination would be among the most critical controls to consider. We have intentionally excluded it here in order to be in some small way more reflective of the \cov{} pandemic which is emerging at the time of conducting this experiment.  

The results of applying this policy are shown in Figure~\ref{fig:FRED-allegheny-county-over-time}, where we show a single simulation of the spread of influenza in Allegheny county over time in both the uncontrolled (left) and controlled (right) cases. Each red dot shows the household of an infectious person.  This policy recommendation can be seen to work as it reduces the maximum number of infected households from 215,799 on day 29 in the uncontrolled case to 65,997 on day 83 in the the uncontrolled case. 
The maximum in the controlled case actually occurs during a second ``outbreak'' after schools re-open (noticeable around day 90 in Figure~\ref{fig:FRED-allegheny-county-over-time-90}). 
While the controlled policy results in two ``spikes,'' our inference procedure accounts for this and correctly controls the maximum number of infectious persons to remain below 10\% of the total at all times. 

\changed{}{
\subsubsection{\cov{} Simulation}
\label{sec:exp:fred:covid}
In this section, we consider \cov{} outbreaks in the Seattle metropolitan area. The controls we consider in these experiments are the following (corresponding to the $\theta$ parameter defined in Table~\ref{tab:covid_prior_params}):
\begin{itemize}
    \item $\theta_1$, a \emph{social distancing}, or the fraction of population practising social distancing.
    \item $\theta_2$, a \emph{workplace closure}, or the fraction of businesses deemed essential and exempt from workplace closure mandates.
    \item $\theta_3$, a \emph{school closure attack rate threshold}, a threshold on fraction of population hospitalized per day. Schools get closed if the daily hospitalization rate exceeds this threshold.
\end{itemize}
Similar to Section~\ref{sec:exp:fred:influenza}, we have put uniform priors on these controllable parameters, as described in Table~\ref{tab:covid_prior_params}.

\begin{figure}
    \centering
    
    \begin{subfigure}[b]{\textwidth}
        \centering
        \includegraphics[]{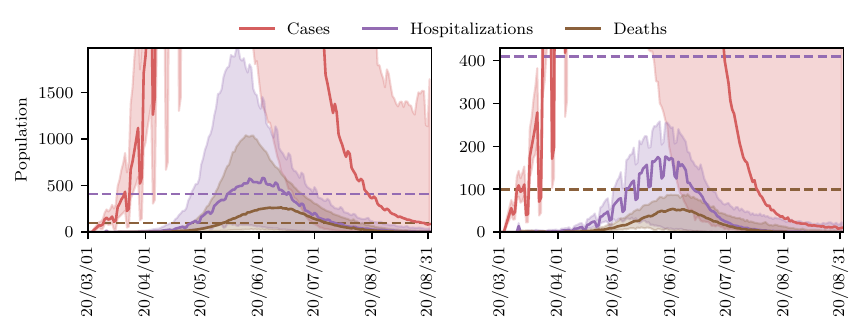}
        \caption{March 2020 - September 2020}
        \label{fig:FRED-seattle-summaries-mar20}
    \end{subfigure}
    \begin{subfigure}[b]{\textwidth}
        \centering
        \includegraphics[]{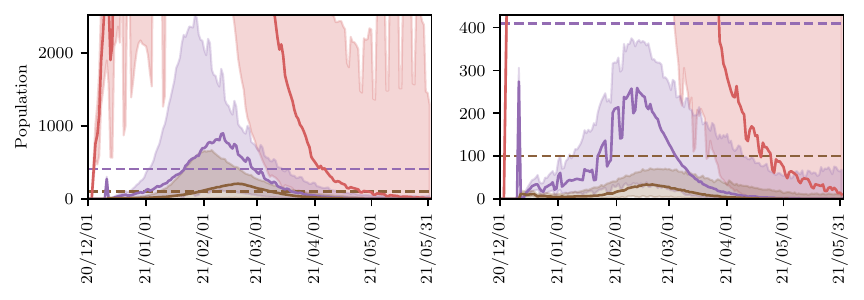}
        \caption{December 2020 - June 2021}
        \label{fig:FRED-seattle-summaries-dec20}
    \end{subfigure}
    \begin{subfigure}[b]{\textwidth}
        \centering
        \includegraphics[]{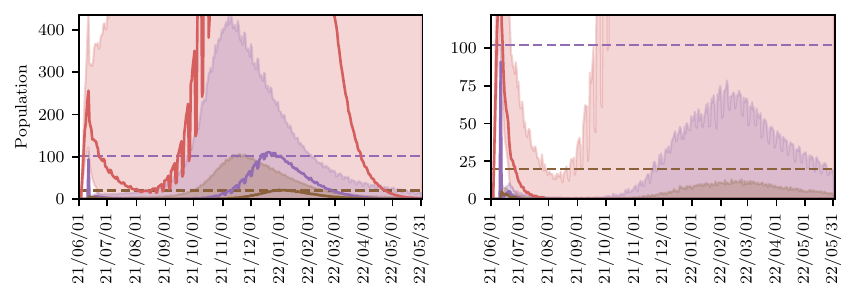}
        \caption{June 2021 - June 2022}
        \label{fig:FRED-seattle-summaries-jun21}
    \end{subfigure}
    \caption{\changed{}{Aggregated statistics from 1,000 FRED simulations of a \cov{} outbreak in the Seattle metropolitan area. We plot the median and confidence intervals between 3\textsuperscript{rd} and 97\textsuperscript{th} percentiles of the \emph{daily} cases, hospitalizations, and deaths. Each row shows a different period of time with different starting conditions and policy goals. The goal in these simulations is to keep the number of daily hospitalizations and deaths below the thresholds specified by dashed lines in matching colors. The left column shows evolution of the outbreak when no control is applied. The right column shows evolution of the outbreak when controls provided by our method are applied. As expected, without control the number of hospitalizations and deaths quickly exceeds the thresholds whereas in the controlled simulations it is ensured that all all the desired conditions are met throughout the simulation period.}}
    \label{fig:FRED-seattle-summaries}
\end{figure}

We put thresholds on the number of daily hospitalizations and daily deaths in different experiments. Hence, $Y_t$ in \eqref{eq:control-inference} is a binary variable indicating if any of the thresholds are exceeded.

\begin{table}[h]  
    \centering
\begin{tabular}{l|l}
     \textbf{Prior} & \textbf{Control} \\
     \hline
      $\theta_1 \sim \, $ \unif{0}{1} & \textit{social distancing}    \\
      $\theta_2  \sim\, $ \unif{0}{1} & \textit{workplace closure}   \\
      $\theta_3  \sim\, $ \unif{1}{$1.5 \times 10^{-4}$} & \textit{school closure attack rate threshold}  \\
     \hline
\end{tabular}
    \caption{Prior over FRED control parameter $\theta = \{\theta_1, \theta_2, \theta_3\}$ for \cov{} simulations.}
    \label{tab:covid_prior_params}
\end{table}

\begin{table}[h]  
    \centering
    \changed{}{
\begin{tabular}{l||l|l|l|l||l|l}
     \multirow{2}{*}{\textbf{Simulation dates}} & \multirow{2}{*}{\textbf{\thead{Active\\infections}}} & \multirow{2}{*}{\textbf{\thead{Total\\hospitalizations}}} & \multirow{2}{*}{\textbf{\thead{Total\\deaths}}} & \multirow{2}{*}{\textbf{\thead{\% Vaccinated}}} &
     \multicolumn{2}{c}{\textbf{Policy goals}}\\\cline{6-7}
     & & & & & Hospitalizations & Deaths\\
     \hline
      1 Mar. 2020 - 1 Sep. 2020 & 400 & 0 & 1 & 0 & 409 & 100\\
      1 Dec. 2020 - 1 Jun. 2021 & 8017 & 2096 & 1459 & 0 & 409 & 100\\
      1 Jun. 2021 - 1 Jun. 2022 & 3024 & 8313 & 2867 & 47 & 102 & 20\\
     \hline
\end{tabular}}
    \caption{\changed{}{Initial conditions and policy goals used in the COVID-19 simulations in Seattle metropolitan area.}}
    \label{tab:covid_initial_params}
\end{table}

\paragraph{Simulator configuration}
Our simulations presented here are on a synthetic population of Seattle metropolitan area incorporated in the FRED simulator (a population size of 3,416,570). The synthetic population closely matches the 2010 census data for the United States with high spatial resolution, differing from the American Community Survey by less than 1\% for large counties. A detailed description of the methodology and further comparisons are provided in \citep{wheaton2009synthesized}.
The parameters of the disease are calibrated to \cov{}, as explained in Section~\ref{sec:models:fred}. We have conducted simulations in three different time periods and the initial conditions of each simulation (number of active infections, total hospitalizations, total deaths and vaccinated people) is set according to the numbers reported by U.S. health officials for the chosen dates (see Table~\ref{tab:covid_initial_params}).

\paragraph{Results}
Figure~\ref{fig:FRED-seattle-summaries} shows the number of daily cases, hospitalizations, and deaths (note that this is not the same as SEIR) for all time periods considered in this experiment. Similar to SEIR plots, it shows results for both the controlled (left) and uncontrolled (right) simulations.

Figure~\ref{fig:FRED-scatter-plot-grid-seattle} shows samples from two-dimensional marginals of the distribution over controlled policy parameters. According to these results, social distancing is the most important control among the three and policy makers should make sure more than 70\% of the population practice it. Although enforcing a more strict school or workplace closure helps as well and allows a slightly wider range for social distancing compliance, they are far less effective than social distancing. Adding more relative control parameters (for example, public mask wearing, recommended vaccine booster shots, restricting large events, imposing travel restrictions, etc.) are planned extensions of future research.

We have conducted \cov{} simulations for three different scenarios:
\begin{itemize}
    \item The 6 months following March 1, 2020 when the outbreak is in its early stages with a few cases.
    \item The 6 months following December 1, 2020 when there is a large number of cases, hospitalizations and deaths.
    \item The 12 months following June 1, 2021 when the number of hospitalizations is the largest among our simulations and there are more than 3000 active infections and near 3000 deaths. However, in this case 47\% of the population are vaccinated at the beginning of the simulation.
\end{itemize}
The goal in the first two scenarios is to avoid the number of daily hospitalizations and daily deaths exceeding 409 and 100 people, respectively, while these numbers are 102 and 20 for the third scenario.
For the third scenario our results in Figure~\ref{fig:FRED-scatter-plot-grid-seattle} show the goals are achieved even with a slightly looser set of controls over the population despite the large number of active infections and hospitalizations and tighter goals. This is indeed the effect of higher vaccinations rates in the initial conditions for the third scenario.

\begin{figure}
    \centering

    \begin{subfigure}[b]{0.49\textwidth}
        \centering
        \includegraphics[width=\textwidth]{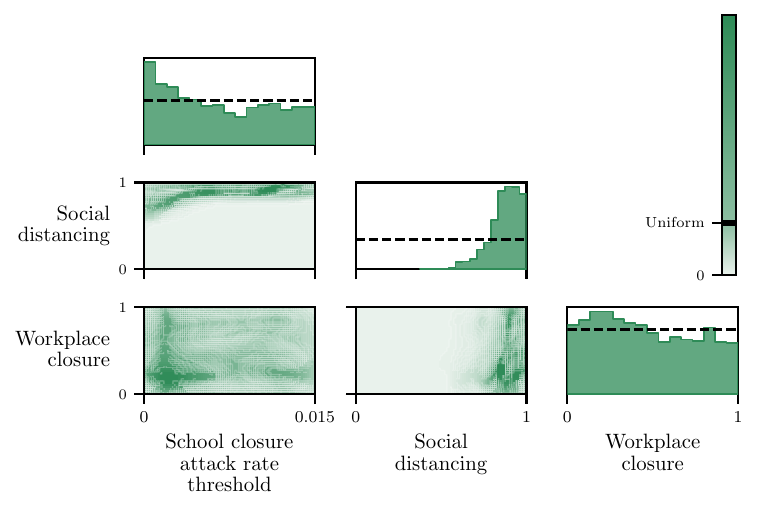}
        \caption{March 2020 - September 2020.}
        \label{fig:FRED-scatter-plot-grid:mar20}
    \end{subfigure}
    \begin{subfigure}[b]{0.49\textwidth}
        \centering
        \includegraphics[width=\textwidth]{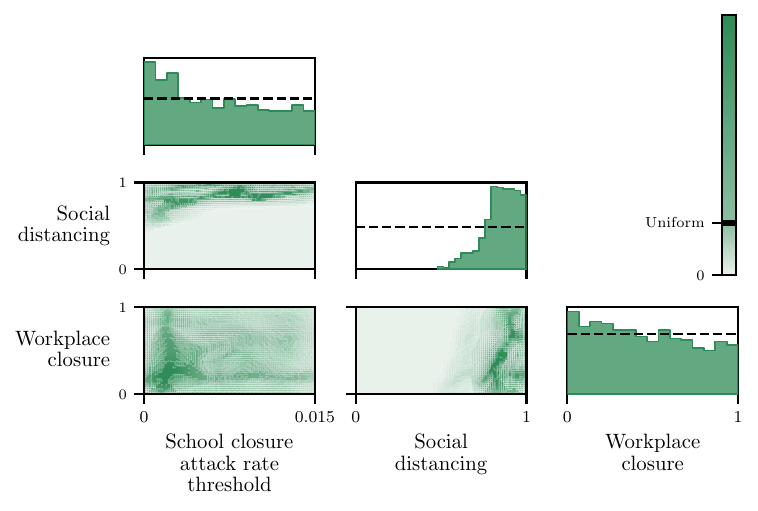}
        \caption{December 2020 - June 2021.}
        \label{fig:FRED-scatter-plot-grid:dec20}
    \end{subfigure}

    \begin{subfigure}[b]{0.5\textwidth}
        \centering
        \includegraphics[width=\textwidth]{figures/FRED/covid/simulation_idx_1_new_vm_plot.pdf}
        \caption{June 2021 - June 2022.}
        \label{fig:FRED-scatter-plot-grid:jun21}
    \end{subfigure}

    \caption{\changed{}{Empirically determined marginals of the full joint distribution over controllable policy parameters that give rise to appropriately controlled outcomes (see Equation \eqref{eq:control-inference}) for simulations of COVID-19 in the Seattle metropolitan area.
    Similar to Figure \ref{fig:FRED-scatter-plot-grid}, the diagonal of each figure shows one-dimensional marginals for each policy parameter, with the uniform prior indicated by the dashed line.
    The remaining array of plots shows kernel density estimation (KDE) plots fitted to samples from the two-dimensional marginals of the joint posterior distribution. For reference, the color corresponding to a uniform prior on the two-dimensional plots is indicated on the color bar on the right.
    Each of the Subfigures \ref{fig:FRED-scatter-plot-grid:mar20}, \ref{fig:FRED-scatter-plot-grid:dec20}, \ref{fig:FRED-scatter-plot-grid:jun21} show the results of simulations for a different time period as indicated. The initial parameters and policy goals for each plot are reported in Table~\ref{tab:covid_initial_params}. We see that among the three policy parameters, social distancing is the most effective in all time periods. Furthermore, putting a more restrictive policy on closing schools of workplaces leads to a wider range of acceptable social distancing policies. Finally, note that in the June 2021 - June 2022 period, when the vaccination rates are high, a looser set of policy enforcement are acceptable even though the desired outcomes for this experiment are much tighter (see Table~\ref{tab:covid_initial_params}).
    }}
    \label{fig:FRED-scatter-plot-grid-seattle}
\end{figure}

Finally, as a part of this project we have created a website\footnote{\url{https://covid19ideas.cs.ubc.ca/}} where users can perform planning as inference in \cov{} simulations for different locations in the U.S. and Canada. It allows users to choose initial conditions of the simulations, policy goals and the set of control parameters (or interventions) to explore. The website automatically runs simulations on the population of the specified location and produces an interactive version of Figures~\ref{fig:FRED-seattle-summaries},\ref{fig:FRED-scatter-plot-grid-seattle}. It also provides the location of infections on the selected location's map, similar to Figure~\ref{fig:FRED-allegheny-county-over-time}. We are working on expanding the features and set of controllable parameters on the website as well.}

\section{Software}
\label{sec:discussion}

In this paper, we have introduced and reviewed control as inference in epidemiological dynamics models and illustrated how this technique can be used to \changed{automate part}{substantially increase the level of automation and precision of some aspects} of policy-making using models at two extremes of expressivity  and  fidelity -- compartmental and agent-based.  We are releasing accompanying source code for our SEIR-type \cov{} model along with bespoke inference code that researchers in the machine learning, control, policy-making, and approximate Bayesian inference communities can use immediately in their own research.  We are also releasing code that demonstrates how a complex, existing, agent-based epidemiological dynamics model can be quickly interfaced to an existing probabilistic programming system so as to automate, even in such a complex \changed{, agent-based epidemiological}{} model, the inference tasks \changed{which are required to be solved when framing}{resulting from the problem formulation of} control as inference.  Again, we stress that the specifics of the models we use to illustrate these techniques are not optimally calibrated for \cov{}, and that we do not report methodological innovation per se, as \changed{}{there are other ways of recombining} existing inference packages \changed{applied to}{and} existing models \changed{can be easily repurposed}{} to \changed{solve the control problem by posing it as inference as we have}{approach the control as inference problem}. \changed{}{In this sense, our principal goal is to demonstrate the feasibility of, and raise interdisciplinary awareness for, such approaches, rather than to champion any specific implementation.}

The first part of the software release accompanying this paper is a pure Python implementation of the \fancyseir{} model adapted to \cov{} from Section~\ref{sec:models:seir}.  It \changed{lives}{is contained} inside an online software version control repository and consists of the model, ``custom'' inference code that implements control as inference  via approximate nested Monte Carlo (NMC), and code that shows how to \changed{do}{perform stochastic} model-predictive control using the model, i.e., the code used to produce Fig.~\ref{fig:exp:seir:mpc}. 

\changed[Update from Epistemix]{The second part of the software release accompanying this paper is a version of the agent-based epidemic simulator FRED \citep{GRE-13-FRED} interfaced to the  probabilistic programming system designed to perform inference in pre-existing simulators, PyProb.  While \cov{}-specific FRED parameters were not available at the time of writing, and only very limited population dynamics data was released publicly (effectively just for Allegheny County, Pennsylvania), we still were able to demonstrate how to do the systems integration required to wire up automated inference in the FRED simulator.  With this integration we were able to explore how policy-makers might search for fine-grained controls using inference, albeit in an influenza model rather than a \cov{} model.  It is our hope that this demonstration will inspire a few things to happen.  One, creating \cov{}-specific parameters and interaction models for FRED would make it, potentially in combination with PyProb, a formidable policy analysis tool, particularly because it could allow very targeted policies to be adapted in efficient and potentially region-specific ways.  Doing this kind of potentially more-precise policy-making could lead to enormous societal benefit, particularly economically.  Indeed, the entire point of this demonstration is to show that with a high-fidelity simulator an expanded and potentially more targeted set of policy interventions can be explored in a, for instance, regionally-specific manner.  As we could only examine a single county in Pennsylvania, we can only hypothesize that the set of effective policy interventions  might differ between counties, states, or countries.  Rapid expansion of the number of counties, provinces, and countries that can be currently simulated in FRED, or any other agent-based model that is subsequently interfaced to PyProb, would allow us or others to quickly test this hypothesis.}{The second part of the software release accompanying this paper is the FRED Modeling Language code for the COVID-19 model executed on the agent-based platform FRED \citep{GRE-13-FRED}, which has been interfaced to the probabilistic programming system PyProb, designed to perform approximate Bayesian inference in pre-existing simulators. The experiments here demonstrate how to achieve the systems integration required to interface automated inference with the FRED simulator.  We believe that model-inference combinations such as FRED+PyProb could provide formidable policy analysis tools with potentially significant societal benefits, particularly because they would allow high-fidelity assessment of region-specific targeted policies.  We expect that the set of effective policy interventions might differ across the regional characteristics of counties, states or countries, as well as their transient epidemiological situations.  Testing this hypothesis depends upon expanding the number of counties, provinces, and countries with simulation profiles in FRED, or in any other agent-based epidemiological model that can be interfaced with a universal and language-agnostic probabilistic programming system such as PyProb.
}   

While the current FRED source code is proprietary, we are happy to discuss the process of PyProb+FRED integration with any interested parties. \changed{}{Furthermore, we have released the PyProb+FRED integration code for our influenza experiment, which uses the older, open-source version of FRED. This code is contained in three separate repositories: a fork of the FRED repository with PyProb integration,\footnote{\url{https://github.com/plai-group/FRED}} a fork of the PyProb code with slight additions required for FRED integration,\footnote{\url{https://github.com/plai-group/pyprob}} and a repository that contains scripts for orchestrating and plotting artifacts from simulation and inference.\footnote{\url{https://github.com/plai-group/covid}}  This last repository also contains a Singularity container \citep{singularity} image bundling all of the necessary software dependencies, including  the repositories above.  This should dramatically reduce the setup time for interested parties, as Singularity images are supported at a large number of existing high performance computing facilities.}

\section{Discussion}

Our experience in conducting this research has led us to identify a number of opportunities for improvement in the 
\changed{}{fields of} simulation-based inference and control \changed{spaces}{}.

\subsection{Software Tools}
Building \changed{and maintaining a}{a simple} SEIR-type model \changed{}{with very few compartments} is a \changed{simple multiple-hour homework-like}{} project \changed{}{of the scope of graduate homework}.  Building and maintaining a simulator like FRED or the US National Large Scale Agent Model \citep{parker2011distributed} is a massive undertaking, \changed{and requires too much time to replicate or, frankly, significantly upgrade}{which would be prohibitive to replicate or significantly extend} in a crisis situation.
As far as we could find \changed{}{with limited search effort} when conducting this work \changed{}{in March 2020}, there \changed{is}{was} neither a central repository of up-to-date open-source agent-based epidemiological models, nor an organizing body \changed{that we could find that}{} we could interface to immediately.
\changed{We may well be wrong, and such a thing does exist; however, if it does, it was insufficiently obvious for us to have found it quickly.}{}

\subsection{Methodology}
\label{sec:methodology}

There appear to be \changed{}{practically very consequential conceptual} gaps between the fields of control, epidemiology, statistics, policy-making, and  probabilistic programming.  To quote \citet{lessler2016trends}, ``the historic separation between the infectious disease dynamics and `traditional' epidemiologic methods is beginning to erode.''  We, like them, see our work as a new opportunity for  ``cross pollination between fields.''  \changed{Again}{As discussed earlier}, the most closely related work that we found in the literature is all focused on automatic model parameter estimation from observational data \citep{kypraios2017tutorial,mckinley2018approximate,toni2009approximate,minter2019approximate,chatzilena2019contemporary}.  These methods and the models to which they have been applied could be repurposed for planning, \changed{in the way we have shown that it can be done}{as we have demonstrated}, simply by changing the \changed{quantities that}{random variables} they \changed{observe}{condition on} to include \changed{control optimality variables as we have demonstrated}{safety or utility metrics}.  \changed{}{Our emphasis therefore lies on the feasibility of planning as inference using existing software tools for approximate Bayesian inference}.

\changed{There are}{Looking closer at the implementation choices, we found} at least two existing papers \changed{that we have identified}{} that explore using probabilistic programming coupled to epidemiological simulators;  \citep{funk2020choices} which used Libbi \citep{murray2013bayesian} and \citep{gram2019hijacking} which used PyProb.  The latter is an example of work that ``hijacks'' a malaria simulator in the same way we ``hijacked'' the FRED simulator in this paper.  Neither explicitly addresses control.  \citep{chatzilena2019contemporary} uses the probabilistic programming system STAN \citep{carpenter2017stan} to address parameter estimation in SEIR-type models, but it too does not explore control, nor could it be repurposed to control an agent-based model with \changed{}{non-differentiable joint densities or, more generally, any external simulator not directly defined in the STAN language}.

Like \citep{kypraios2017tutorial,mckinley2018approximate,toni2009approximate,minter2019approximate,chatzilena2019contemporary} we too could have demonstrated automated parameter estimation in both of the models that we consider, for instance to automatically adjust uncertainty in disease-specific parameters by conditioning on observable quantities \changed{}{such as death counts} that are measured \changed{as, for instance, \cov{} is evolving}{online during a breakout}. \changed{For instance, we could also condition the model on total reported number of deaths due to \cov{}, collected on a daily or weekly interval.}{}  However, as the epidemiological community already relies upon long-standing methods established for estimating confidence intervals for model parameters during evolving pandemics, we exclusively restricted ourselves to demonstrating how to achieve control via inference, assuming that \changed{all control inference tasks are conducted using}{the disease parameter} priors \changed{that are posteriors from other,}{have been obtained from} established parameter estimation techniques.  Combining the two kinds of observations in a single inference task is technically straightforward but does require care in interpretation.


\section{Final Thoughts}
\label{sec:final-thoughts}

\changed{}{At the time of writing all authors, apart from J.G. and D.C. at Epistemix Inc., Pittsburgh, were principally affiliated with} the ``Programming Languages for Artificial Intelligence'' (PLAI) research group at UBC, Vancouver, \changed{the research affiliation of all the authors,}{which} is primarily involved in developing next-generation \changed{}{probabilistic} AI tools and techniques. We felt, however, that \changed{current}{the acute} circumstances demanded we lend whatever we could to the global fight against \cov{}. Beyond the specific contributions outlined above, our secondary aim in writing this paper was to encourage other researchers to contribute their expertise to the fight against \cov{} as well. We believe the world will be in a better place more quickly if they do. \changed{and hope we have contributed to bringing this about.}{}

\section{Acknowledgements}
\label{sec:acknowledgements}

\changed{We acknowledge the support of the Natural Sciences and Engineering Research Council of Canada (NSERC), the Canada CIFAR AI Chairs Program, Compute Canada, Intel, and DARPA under its D3M and LWLL programs.}{We acknowledge the support of the Natural Sciences and Engineering Research Council of Canada (NSERC), the Canada CIFAR AI Chairs Program, and the Intel Parallel Computing Centers program. This material is based upon work supported by the United States Air Force Research Laboratory (AFRL) under the Defense Advanced Research Projects Agency (DARPA) Data Driven Discovery Models (D3M) program (Contract No. FA8750-19-2-0222). Additional support was provided by UBC's Composites Research Network (CRN), Data Science Institute (DSI), Support for Teams to Advance Interdisciplinary Research (STAIR) Grants, the Innovation for Defence Excellence and Security (IDEaS) program, under its COVID-19 Challenges (CPCA-0397), and a CIFAR AI and COVID-19 Catalyst Grant. This research was enabled in part by technical support and computational resources provided by WestGrid (https://www.westgrid.ca/) and Compute Canada (www.computecanada.ca).}

\clearpage
\bibliographystyle{unsrtnat}  
\bibliography{references}

%














\end{document}